%% file: resumhex_final.tex
\documentclass[12pt]{article}

\pdfoutput=1
\usepackage{color}
\usepackage{epsfig, palatino}
\usepackage{pstricks,pst-node,pst-tree}
\usepackage{epic}
\usepackage{mathrsfs}
\usepackage{ae} 
\usepackage[T1]{fontenc}
\usepackage[ansinew]{inputenc}
\usepackage{amsmath}
\usepackage{amssymb}
\usepackage{graphicx}
\usepackage{ulem}
\usepackage{color}
\definecolor{darkblue}{cmyk}{0.9,0.9,0,0}
\usepackage[colorlinks=true,linkcolor=darkblue,citecolor=darkblue,urlcolor=darkblue]{hyperref}
\usepackage{cite}
\usepackage{hyperref}
\usepackage{wasysym}
\usepackage{varioref}
\usepackage{makeidx}
\usepackage[english]{babel}
\usepackage{simplewick}
\usepackage{array}

\newcommand{\comment}[1]{}

\newcommand{\beq}{\begin{equation}}
\newcommand{\eeq}{\end{equation}}
\newcommand{\beqq}{\begin{equation*}}
\newcommand{\eeqq}{\end{equation*}}
\newcommand\beqa{\begin{eqnarray}}
\newcommand\eeqa{\end{eqnarray}}
\newcommand\beqaa{\begin{eqnarray*}}
\newcommand\eeqaa{\end{eqnarray*}}
\newcommand\bea{\begin{array}}
\newcommand\eea{\end{array}}

\newcommand{\nn}{\nonumber}

\newcommand{\neqa}{\nonumber\end{eqnarray}} 
\newcommand{\la}[1]{\label{#1}}

\newcommand{\ga}{\mathfrak{a}}
\newcommand{\gb}{\mathfrak{b}}

\renewcommand{\d}{\partial}

\newcommand{\<}{{\langle}}
\renewcommand{\>}{{\rangle}}

\newcommand{\cA}{{\cal A}}

\newcommand{\cC}{{\cal C}}

\newcommand{\re}{\relax{\rm I\kern-.18em R}}

\newcommand{\ch}{{\rm cosh}}

\renewcommand{\sp}{p\hspace{-.40em}/}

\definecolor{darkgreen}{rgb}{0.0, 0.45, 0.0}

\def\XXint#1#2#3{{\setbox0=\hbox{$#1{#2#3}{\int}$}
\vcenter{\hbox{$#2#3$}}\kern-.5\wd0}}

\def\su2{{SU(2)}}

\def\[{\left[}
\def\]{\right]}

\def\({\left(}
\def\){\right)}
\def\[{\left[}
\def\]{\right]}

\def\<{\langle}
\def\>{\rangle}

\def\i2{\frac{i}{2}}

\def\spi{\relax{\rm \pi\kern-0.5em /}}
\def\sA{\relax{\rm A\kern-0.5em /}}
\def\sp{\relax{\rm p\kern-0.5em /}}
\def\sd{\relax{\rm \d\kern-0.5em /}}
\def\sk{\relax{\rm k\kern-0.5em /}}
\def\sn{\relax{\rm n\kern-0.5em /}}
\def\sl{\relax{\rm l\kern-0.5em /}}
\def\sP{\relax{\rm P\kern-0.7em /}}
\def\sBethe{\relax{\rm \Bethe\kern-0.5em /}}

\def\cC{{\cal C}}

\def\cR{{\cal R}}
\def\cO{{\cal O}}

\def\cP{{\cal P}}

\def\cW{{\cal W}}

\def\2F1{\,_2{\rm F}_1}

\makeindex

\begin{document}

\thispagestyle{empty}

\renewcommand{\thefootnote}{\fnsymbol{footnote}}
\setcounter{page}{1}
\setcounter{footnote}{0}
\setcounter{figure}{0}

%%%%%%%%%%%%%%%%%%%%%%%%%%%%%%%%%%%%%%%%%%%%%%%%%%%%%%%%%%%%%%%%%%%%%%%%%%%%%%%%%%%%%%%%%%%%%%%%%
\begin{center}

$$$$

{\Large\textbf{\mathversion{bold}Hexagon POPE: effective particles and \\
tree level resummation
}}

\vspace{1.0cm}

\textrm{Luc\'ia C\'ordova$^{{\footnotesize\pentagon},{\footnotesize\hexagon}}$}
\\ \vspace{1.2cm}
\footnotesize{\textit{
$^{\pentagon}$Perimeter Institute for Theoretical Physics,
Waterloo, Ontario N2L 2Y5, Canada\\
$^{\hexagon}$Department of Physics and Astronomy \& Guelph-Waterloo Physics Institute, University of Waterloo, Waterloo, Ontario N2L 3G1, Canada\\
}  
\vspace{4mm}
}

\par\vspace{1.5cm}

\textbf{Abstract}\vspace{2mm}
\end{center}

We present the resummation of the full Pentagon Operator Product Expansion series of the hexagon Wilson loop in planar $\mathcal N=4$ SYM at tree level. We do so by considering the one effective particle states formed by a fundamental flux tube excitation and an arbitrary number of the so called \textit{small fermions} which are then integrated out. We present our proposals for the one effective particle measures at finite coupling. By evaluating these measures at tree level and summing over all one effective particle states we reproduce the full 6 point tree level amplitude.
\noindent

\setcounter{page}{1}
\renewcommand{\thefootnote}{\arabic{footnote}}
\setcounter{footnote}{0}

 \def\nref#1{{(\ref{#1})}}

\newpage

\tableofcontents

\parskip 5pt plus 1pt   \jot = 1.5ex
\newpage
%%%%%%%%%%%%%%%%%%%%%%%%%%%%%%%%%%%%%%%%%%%%%%%%%%%%%%%%%%%%%%%%%%%%%%%%%%%%%%%%%%%%%%%%%%%%%%%%%
\section{Introduction}

$\mathcal N=4$ Super Yang Mills is a special theory in which the integrability of the planar regime allows us to compute observables at any value of the coupling. Two such observables are the expectation value of null polygonal Wilson loops and scattering amplitudes, which are dual to each other \cite{AM,AmplitudeWilson}. 

These observables are what the Pentagon Operator Product Expansion (POPE) \cite{short} program  studies. This is an expansion around the collinear limit in which the Wilson loop expectation value is given as an infinite sum over flux tube excitations created at the bottom and absorbed at the top of the polygon. The building blocks have been bootstrapped at any value of the coupling and matched against data \cite{data,2pt,fusion,Belitsky:2014sla,Andrei1,Andrei2,FrankToAppear,shortSuper,FF} and in \cite{shortHexagon}, the complete POPE series for the hexagon was unveiled.

A natural question to ask is if this expansion can be resummed to reproduce the full kinematical dependence of the amplitude. In general this is not a simple problem since already at tree level we need to sum over an infinite set of excitations. The resummation of the POPE was considered before both at weak and strong coupling. At weak coupling, in \cite{georgiosjames,georgios} a procedure for the resummation of the single particle gluon bound states or double scaling limit was presented whereas at strong coupling in \cite{Fioravanti:2015dma,Bonini:2015lfr} the contribution of gluons and mesons were studied.

However, the POPE weak coupling resummation where the full set of flux tube excitations is taken into account is still pending. This might seem a rather difficult task since we need to sum over the contributions of all possible combinations of gluons, scalars and fermions. The way out is that we can reorganize the excitations into effective particles. As was put forward in \cite{FF} the number of effective particles needed to reproduce an amplitude grows very slowly with loop order. In fact, to compute the six point amplitude at tree level and one loop only states with one effective particle are needed. 

The one effective particle states are formed by one fundamental excitation --could be a gluon bound state, a scalar or a large fermion-- and an arbitrary number of small fermions (antifermions) which are then integrated out. As we shall see, the fundamental excitation and the small fermions are organized in Bethe strings which allows us to perform the integrations straightforwardly. Since the string patterns are derived from the matrix part of the POPE integrand which is coupling independent, we use these results to compute the one effective particle measures at finite coupling. We later evaluate these measures at tree level and sum over all one effective particle states to get a rational function which reproduces the 6 point tree level amplitude for \textit{general kinematics}. In the end, the tree level resummation turns out to be very simple.

The paper is organized as follows. In section \ref{Hexagon OPE} we first review the hexagon POPE building blocks and consider the one effective particle states, presenting their measures at finite coupling. In section \ref{tree resum} we evaluate these measures at tree level and perform the resummation. In the same section we explain how this result reproduces the NMHV 6 point amplitude at tree level. We conclude with some final remarks.

\section{Hexagon POPE and one effective particle states}\la{Hexagon OPE}

\begin{figure}
\centering
\includegraphics{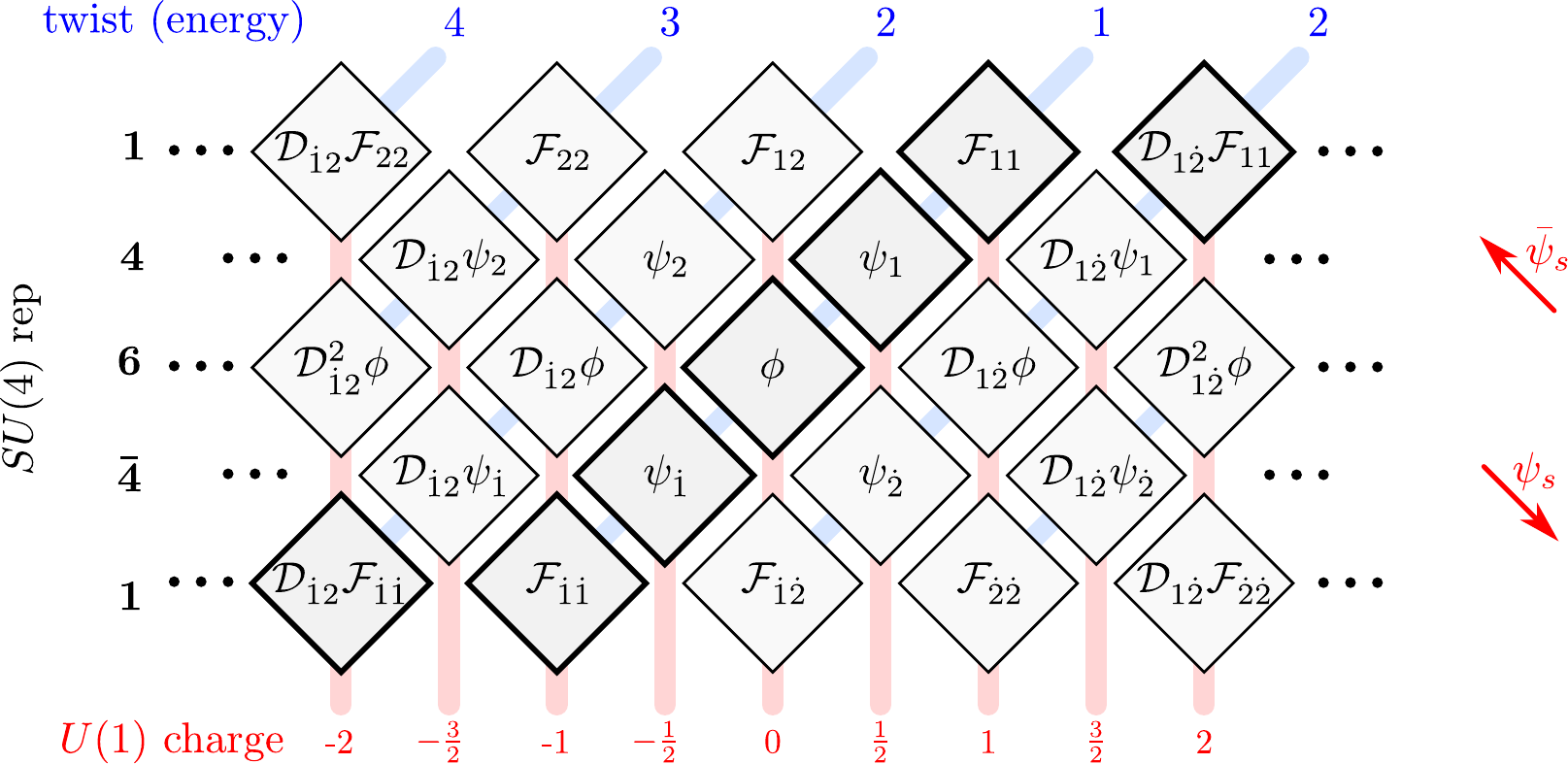}
\caption{Table of effective weak coupling excitations presented in \cite{2pt,FF}. The fundamental excitations are in the bold squares. The effective particles can be formed by adding small fermions or antifermions.}
\label{tableWEAK}
\end{figure}

Let us first recall the results spelled out in \cite{shortHexagon} that provide the building blocks for our derivation. The hexagon Wilson loop we want to compute is given as a sum over all possible flux tube excitations parametrized by rapidities $u_i$
\beq
\mathcal{W}_6=
\vcenter{\hbox{\includegraphics[scale=.13]{hexagon.pdf}}}
=\sum_m\frac{1}{S_m}\int\frac{du_1\ldots du_m}{(2\pi)^m}\,\Pi_{\text{dyn}}\times\Pi_{\text{FF}}\times\Pi_{\text{mat}}\la{gralhex}\,,
\eeq
where $S_m$ is a symmetry factor. The integrand is nicely factored out into a \textit{dynamical} and \textit{form factor} parts which carry the coupling dependence and a \textit{matrix} factor which takes care of the R-symmetry structure of the theory. The first factor is universal and reads
\beq
\Pi_{\text{dyn}}=\prod_i \mu(u_i)e^{-E(u_i)\tau + ip(u_i)\sigma +im_i\phi}\times\prod_{i<j}\frac{1}{|P(u_i|u_j)|^2}\la{dyn}\,,
\eeq
where $\{\tau,\sigma,\phi\}$ are respectively the flux tube time, space and angle (related to the three conformal cross ratios of the hexagon); $E(u_i)$, $p(u_i)$ and $m_i$ are the energy, momentum and angular momentum of the excitation; $P(u_i|u_j)$ are the pentagon transitions between different excitations and $\mu(u_i)$ the corresponding measures. We will often use the notation $\hat\mu(u)=\mu(u)e^{-E(u)\tau + ip(u)\sigma +im\phi}$. The fundamental flux tube excitations are gluon bound states, fermion, antifermion and scalar: $\{F_b,\psi,\bar\psi,\phi\}$. They are represented in the bold squares of figure~\ref{tableWEAK}.

The next factor in the POPE integrand is non trivial only for Next to Maximally Helicity Violating (NMHV) hexagons (i.e. charged pentagons) and is given by
\beq
\Pi_{\text{FF}}=g^{\tfrac{1}{8}r_1(r_1-4)+\tfrac{1}{8}r_2(r_2-4)}\times\prod_i h(u_i)^{r_2-r_1}\la{ff}\,,
\eeq
where $r_i$ is the R-charge in the $i$-th pentagon and $h(u_i)$ are the so called \textit{form factors} derived in \cite{FF}. The last factor is the matrix part which takes into account the contraction of the SU(4) R-symmetry indices of each pentagon. It has the following form \cite{FrankToAppear}
\beqa
\Pi_{\text{mat}}&=&\frac{1}{K_1!K_2!K_3!}\int\prod\limits_{i=1}^{K_1}\frac{dw_i^1}{2\pi}\prod\limits_{i=1}^{K_2}\frac{dw_i^2}{2\pi}\prod\limits_{i=1}^{K_3}\frac{dw_i^3}{2\pi}\times\nn\\
&&\times\frac{g({\bf w}^1)g({\bf w}^2)g({\bf w}^3)}{f({\bf w}^1,{\bf w}^2)f({\bf w}^2,{\bf w}^3)f({\bf w}^1,{\bf v})f({\bf w}^2,{\bf s})f({\bf w}^3,{\bf \bar v})}\la{mat}\,,
\eeqa
where $w_i$ are auxiliary roots of three different types and $\{v_i,s_i,\bar v_i\}$ are rapidities for fermions, scalars and antifermions, respectively; the functions $g({\bf w})=\prod_{i<j}(w_i-w_j)^2[(w_i-w_j)^2+1]$, $f({\bf w},{\bf v})=\prod_{i,j}[(w_i-v_j)^2+\frac{1}{4}]$ and the number of auxiliary rapidities $K_j$ are the solution to the equations
\beqa
N_\psi -2 K_1 +K_2  &=& \delta_{r_1,3}\,,\nn\\
N_\phi + K_1- 2 K_2 + K_3 &=& \delta_{r_1,2}\,,\la{Ks}\\
N_{\bar\psi}\,\,\, + K_2 -2K_3 &=& \delta_{r_1,1}\nn\,,
\eeqa
where $N_\psi$, $N_\phi$ and $N_{\bar\psi}$ are respectively the number of fermions, scalars and antifermions.

Together with the pentagon transitions, form factors and energies presented in \cite{data,FF}, these expressions are all the necessary ingredients to compute the hexagon Wilson loop as a series in the collinear limit. From them we shall derive the one effective particle measures that we later resum at tree level. 

As we shall be working mostly with NHMV amplitudes, let us review some useful notation. The hexagon super Wilson loop can be decomposed into POPE components $P^{[r_1]}P^{[r_2]}$, where $r_{i}$ is the total R-charge in the $i$-th pentagon and it takes values $0\leq r_i\leq4$. For N$^k$MHV components we have that $\sum_ir_i=4k$. Therefore the hexagon NMHV ($r_1+r_2=4$) has five different POPE components. Depending on which POPE component we are considering, there is a subset of allowed excitations  determined by the representation of the SU(4) R-symmetry in which the state transforms. For instance, for the NMHV component $P^{[2]}P^{[2]}$ we could have the excitations: $\phi,\,\bar\psi\bar\psi,\,\psi\psi F_{a}$, etc.

\begin{figure}
\centering
\includegraphics[scale=.7]{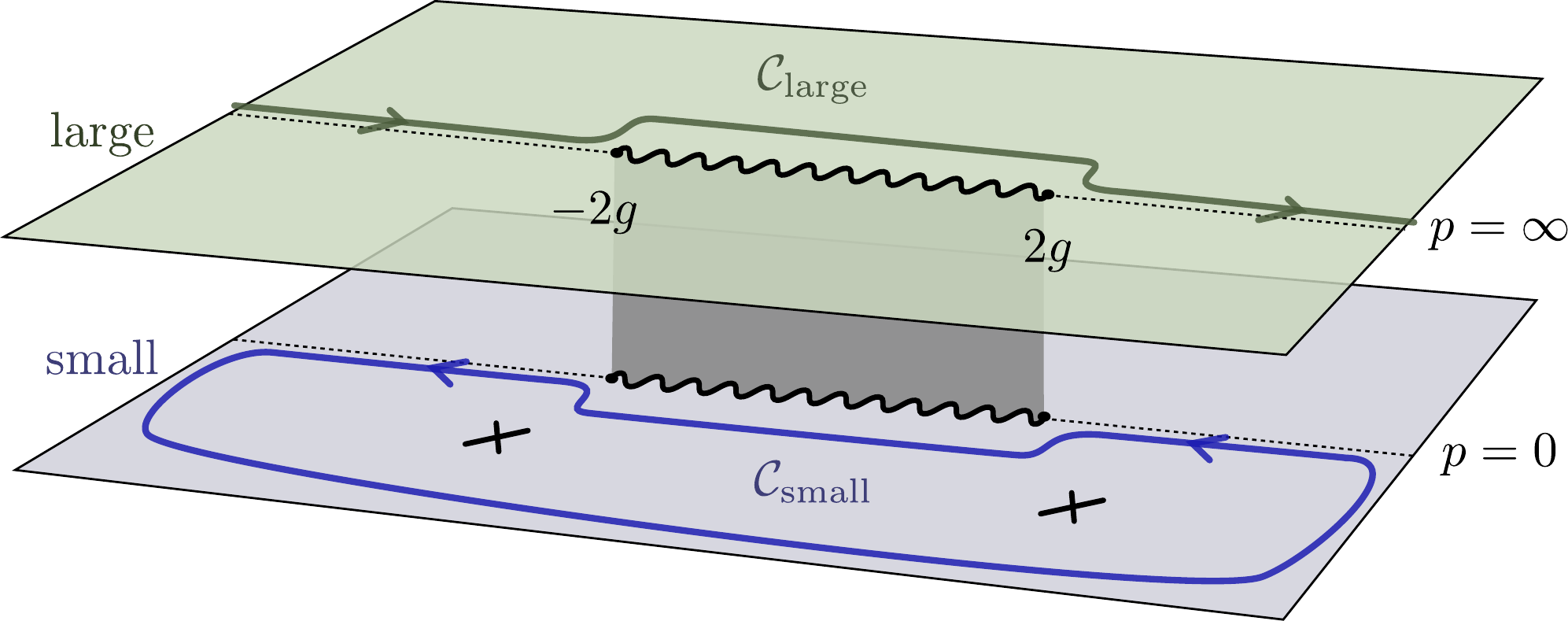}
\caption{The fermion contour of integration in terms of the rapidity $u$ involves a Riemann surface with two sheets --one in which the momenta is large and another one in which it is small-- connected by a branch cut between $u=-2g$ and $u=2g$. The original contour in \cite{2pt} can be split into two different contours $\mathcal C_{\text{large}}$ and $\mathcal C_{\text{small}}$, each one on a different sheet. The small fermion contour $\mathcal C_{\text{small}}$ might enclose poles coming from interactions with other excitations.}
\label{fermion_contour}
\end{figure}
The fermionic excitations have the important feature that they can be separated into \textit{large} and \textit{small fermions}. This is because in terms of the Bethe rapidity the fermion integration contour involves two different Riemann sheets, one in which the fermions have large momenta and another one in which their momenta is small \cite{2pt}. The integration contour can be split into two so that each new contour lives only in one of the two Riemann sheets, as figure~\ref{fermion_contour} shows. 

\begin{figure}
\centering
\includegraphics{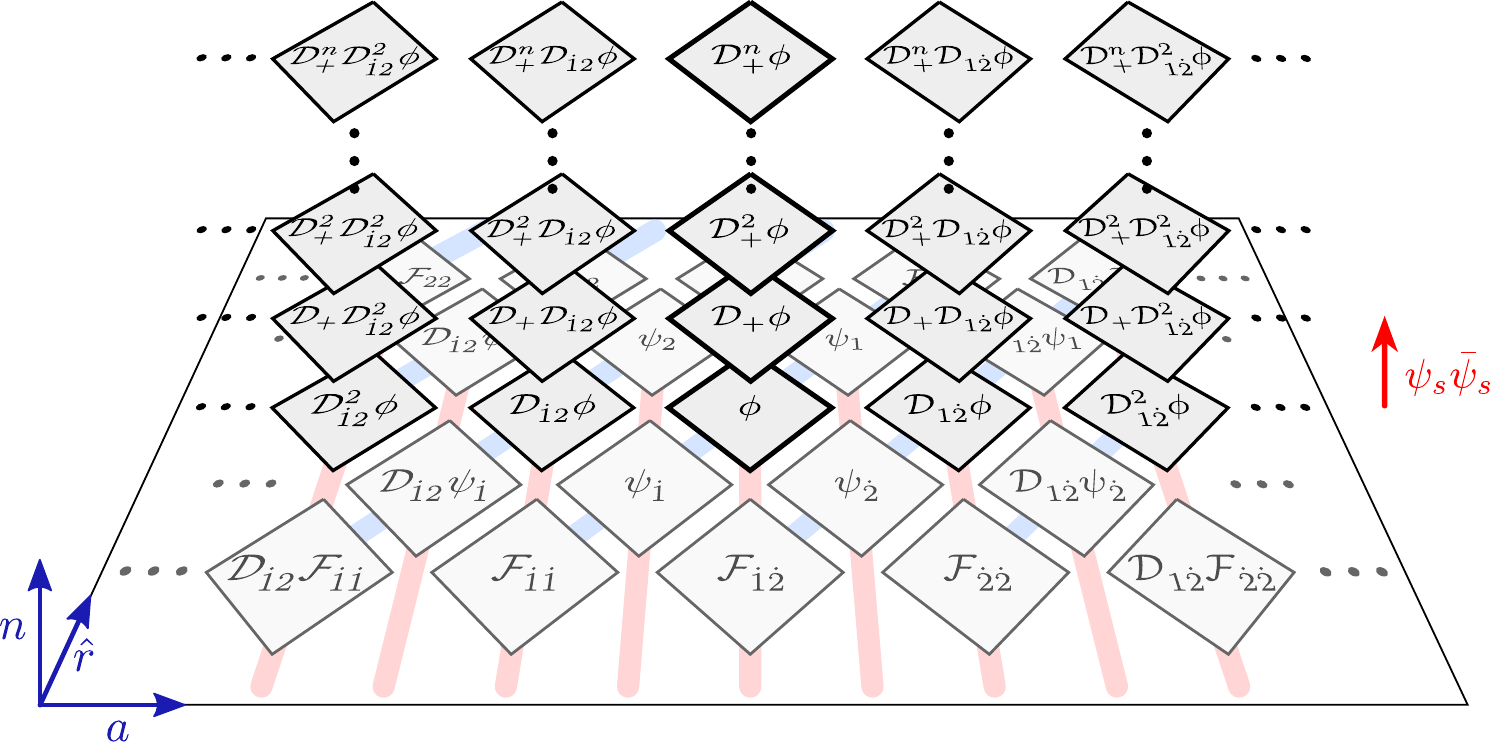}
\caption{Table of effective weak coupling excitations including the first $n$ descendants of the particles transforming in the vector representation of SU(4). The plane in the bottom contains the primary excitations depicted in figure~\ref{tableWEAK}. A descendant is formed by acting with a pair $\psi_s\bar\psi_s$ (or derivative $D_+$) on one of these excitations. Moving away from this plane in the vertical direction corresponds to adding more descendants. An effective excitation is characterized by its helicity $a$, the SU(4) R-symmetry representation labelled by $\hat r$ and its number of descendants $n$.}
\label{descendants}
\end{figure}

In the small sheet there are potential poles enclosed by the contour $\mathcal C_{\text{small}}$. These would come from the interaction of the small fermions with other excitations. When attached to another particle, the small fermions $\psi_s$ act as supersymmetry generators \cite{AldayMaldacena} and create a sea of effective excitations, some of which are shown in figure~\ref{tableWEAK}. We can also add an arbitrary number of pairs of small fermion-antifermion $\psi_s\bar\psi_s$ (or derivatives $D_+$), creating the so called \textit{descendants} depicted in figure~\ref{descendants}. The name is because, as explained in \cite{Ben,Gaiotto:2010fk,Gaiotto:2011dt}, at weak coupling there is an enhancement of symmetry from SU(4) to SL(2|4) and the flux tube excitations can be packed in SL(2) conformal blocks. The primaries correspond to the excitations in the plane presented in figure~\ref{tableWEAK}, obtained by the action of small fermions or antifermions. On the other hand, the descendants correspond to the excitations in the vertical direction in figure~\ref{descendants} obtained by the action of pairs of small fermion-antifermion $\psi_s\bar\psi_s$. Although this symmetry is exact only up to one loop, we will keep the terminology for the finite coupling discussion.

In sum, we can distinguish an effective excitation by its position on the three dimensional space shown in figure~\ref{descendants}. The three parameters are: the helicity of the excitation $a$, its number of descendants $n$ and the SU(4) R-symmetry representation in which it transforms $\hat r$ (distinguishing between the two sets of singlet excitations shown in the first and last row of figure~\ref{tableWEAK}). As we explain in the following sections, to perform the tree level resummation we fix the SU(4) representation and sum the measures of the effective particles over $a$ (from $-\infty$ to $\infty$) and over $n$ (from 0 to $\infty$).
\subsection*{One effective particle states}

A one effective particle $\Phi$ is formed by one fundamental excitation --referred to in the following as $\Phi_0$-- and an arbitrary number $N_{\psi_s}\,(N_{\bar\psi_s})$ of small fermions (antifermions) that are integrated out. Higher number of effective particles include more than one fundamental excitation. In terms of effective excitations, a POPE hexagon component reads
\beq
P^{[r_1]}P^{[r_2]}=\sum\limits_{\Phi}\,\int\frac{du}{2\pi}\,e^{-E_{\Phi}(u)\tau+ip_{\Phi}(u)\sigma+im_{\Phi}\phi}\mu^{[r_1,r_2]}_{\Phi}(u)+\ldots\la{sum1effpt}\,,
\eeq
where the dots account for higher effective particles and we have adopted the notation $\mu_\Phi^{[r_1,r_2]}(u)\equiv\,\left[g^{\frac{1}{8}r_1(r_1-4)+\frac{1}{8}r_2(r_2-4)} h_{\Phi}(u)^{r_2-r_1}\right]\mu_\Phi(u)$. The effective measures $\mu_\Phi$ will be given by an expression of the sort
\beq
\mu_\Phi=\int\limits_{\cC_{\text{small}}}\frac{dv_1d\bar v_1\dots}{(2\pi)^{N_{\psi_s}+N_{\bar\psi_s}}}\Pi_{\text{dyn}}(\Phi_0\psi_s^{N_{\psi_s}}\bar\psi_s^{N_{\bar\psi_s}})\,\Pi_{\text{FF}}(\Phi_0\psi_s^{N_{\psi_s}}\bar\psi_s^{N_{\bar\psi_s}})\,\Pi_{\text{mat}}(\Phi_0\psi_s^{N_{\psi_s}}\bar\psi_s^{N_{\bar\psi_s}})\la{effmu}\,,
\eeq
where $v_i\,(\bar v_i)$ are the rapidities of the small fermions (antifermions) and $\cC_{\text{small}}$ is the small fermion contour shown in figure~\ref{fermion_contour}.

One advantage of this approach is that the amount of effective particles needed to fully reproduce a scattering amplitude at a given perturbative order grows very slowly with the loop order. For instance, one effective particle states are sufficient to reproduce amplitudes at tree level and one loop, states with two effective particles are enough up to five (four) loops for MHV (NMHV) amplitudes, etc \cite{FF}. Moreover, having a compact formula for effective particles with arbitrary number $N_{\psi_s}\,(N_{\bar\psi_s})$ of small fermions represents a huge simplification for the starting point of the resummation. In the following we describe the combinatorics involved in the small fermion integrations, the reader might want to skip this discussion and jump to the next section.

The small fermion integrations in \eqref{effmu} can be carried out straightforwardly by residues. The relevant poles between different rapidities arise from the matrix part\footnote{Here we redefine the pentagon transition between small fermion and gluon bound state in the following way $P_{F_b|\psi_s}(u|v)_{\text{here}}=(u-v+ia/2)^{-1}P_{F_b|\psi_s}(u|v)_{\text{\cite{FF}}}$ so that the factor $(u-v+ia/2)^{-1}$ is part of the matrix part and the statement is indeed true for all flux tube excitations.}. Since this part of the integrand is coupling independent, the structure of poles will be the same at any value of the coupling. Although taking residues might be trivial, we need to do so for an arbitrary number of integration variables (remember we can add infinite pairs $\psi_s\bar\psi_s$). As we explain in the following, instead of taking all the possible residues we can find a pattern in which the small fermions attach to the fundamental excitation forming a Bethe string. Then we would only need to multiply by an appropriate combinatoric factor. Computing the integrals in this way is much more efficient and in practice it is the only way to account for a very large number of small fermions and auxiliary rapidities.

Let us explain how the structure of these strings arises with a simple example. Consider a scalar excitation and its descendants $\phi(\psi_s\bar\psi_s)^n$ which contribute to the POPE component $P^{[2]}P^{[2]}$. These are the excitations in the tower at the center of figure~\ref{descendants}. We want to find the pattern in which the small fermions and antifermions attach to the scalar. 

For $n=0$ the matrix part is trivial so that we have only the scalar measure $\mu_\phi(u)$. For $n=1$, the effective measure (multiplied by the corresponding square propagation factor) is given by
\beq
\hat\mu_{\phi\psi_s\bar\psi_s}(u)=\int\limits_{\cC_{\text{small}}}\frac{dv_1\,d\bar v_1}{(2\pi)^2}\frac{\hat\mu_\phi(u)\hat\mu_{\psi_s}(v_1)\hat\mu_{\bar\psi_s}(\bar v_1)}{|P_{\phi|\psi_s}(u|v_1)|^2\,|P_{\phi|\bar\psi_s}(u|\bar v_1)|^2\,|P_{\psi_s|\bar\psi_s}(v_1|\bar v_1)|^2}\times\frac{1}{g}\times\Pi_{\text{mat}}(\phi\psi_s\bar\psi_s)\la{mu n1}\,,
\eeq
where $1/g$ is the form factor contribution. Since the matrix part is what determines the poles, let us write it explicitly. According to~\eqref{Ks}, we have one auxiliary root of each type which leads to 
\beqa
\Pi_{\text{mat}}(\phi\psi_s\bar\psi_s)&=&\int_\mathbb{R}\frac{dw^1_1dw^2_1dw^3_1}{(2\pi)^3}\frac{1}{f(w^1_1,w^2_1)f(w^2_1,w^3_1)f(w^1_1,v_1)f(w^2_1,u)f(w^3_1,\bar v_1)}\,,\la{intmuphimat}\\
&=&\frac{6 v_1^2+4 u^2+6 \bar{v}_1^2-4 u\bar{v}_1-4 u v_1-8 v_1 \bar{v}_1+45}{\left(\left(u-v_1\right){}^2+\frac{9}{4}\right) \left(\left(u-\bar{v}_1\right){}^2+\frac{9}{4}\right)\(\left(v_1-\bar{v}_1\right){}^2+4\)}\,,\la{muphimat}
\eeqa
where $f(u,v)=(u-v)^2+1/4$ as before. Next we would replace this factor in \eqref{mu n1} and integrate over $v_1,\,\bar v_1$. The integration contour $\cC_{\text{small}}$ is the half moon in the lower half of the complex plane shown in figure~\ref{fermion_contour}. This means that the poles that we are going to pick are $v_1=u-3/2i$ and $\bar v_1=u-3/2i$. In other words, we find that the Bethe string is formed by a scalar with rapidity $u$ and a small fermion and antifermion both attached at a distance $3/2i$ below it. As we shall see, it is convenient to think of the small fermions as auxiliary roots and obtain the same string skipping the intermediate step \eqref{muphimat}. In this way we can perform all the integrations in \eqref{mu n1} (over small fermions and auxiliary roots) by studying the structure of the poles in the matrix part \eqref{intmuphimat} and finding a pattern in which we take the residues. 

Let us see how we find the same string in this manner. Each function $f(v,u)$ in \eqref{intmuphimat} gives two poles: one at $v=u-i/2$ and another one at $v=u+i/2$. Since we know that the small fermion rapidities should be evaluated in the lower half of the complex plane, we shall take the residues at the poles with negative imaginary part. Starting with the auxiliary root $w^2_1$ we take the residue at $w^2_1=u-i/2$. After that, the denominator in \eqref{intmuphimat} becomes $f(w^1_1,u-i/2)f(u-i/2,w^3_1)f(w^1_1,v_1)f(w^3_1,\bar v_1)$ so next we take the residues at $w^1_1=u-i$ and $w^3_1=u-i$ (here the order does not matter). That leaves us with the product $f(u-i,v_1)f(u-i,\bar v_1)$ and the residues at $v_1=u-i3/2$, $\bar v_1=u-i3/2$ which give us the same string as before. This pattern in which we take the residues can be represented by the following picture
\begin{center}
\includegraphics{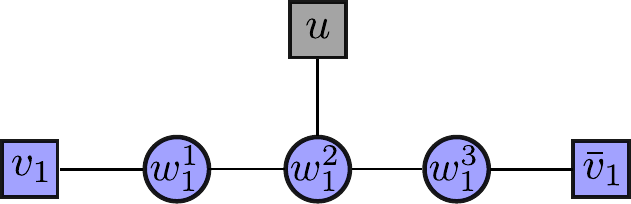}
\end{center}
where the top node in gray corresponds to the scalar with rapidity $u$, the square node in the left (right) represents the small fermion (antifermion) and the circular nodes the auxiliary roots\footnote{Note that in the previous pattern we can identify the line with the blue nodes with the Dynkin diagram of SL(2|4).}. We start by integrating out the node closer to the fundamental excitation, in this case~$w^2_1$. The residues are taken at positions $u-i\#/2$, where $\#$ is the number of line segments between the fundamental node and the one we are integrating out.

\begin{figure}[t]
\centering
\def\svgwidth{15cm} \input{stringsP2P2.pdf_tex}
\caption{String patterns in which the small fermions attach for excitations in the vector representation of SU(4). These are the excitations appearing in the POPE component $P^{[2]}P^{[2]}$. The box at the top of each column labels the fundamental excitation to which the small fermions attach. In black, the fundamental excitation with rapidity $u$; in gray, the small fermions (antifermions) needed to have an excitation transforming in this representation. In blue (green) the first (second) pair of $n$ descendants ($\psi_s\bar\psi_s$). The arrows show the separation between the different excitations in the rapidity plane. As we can see, the patterns are symmetric between positive and negative helicity states.}
\label{stringsP2P2}
\end{figure}
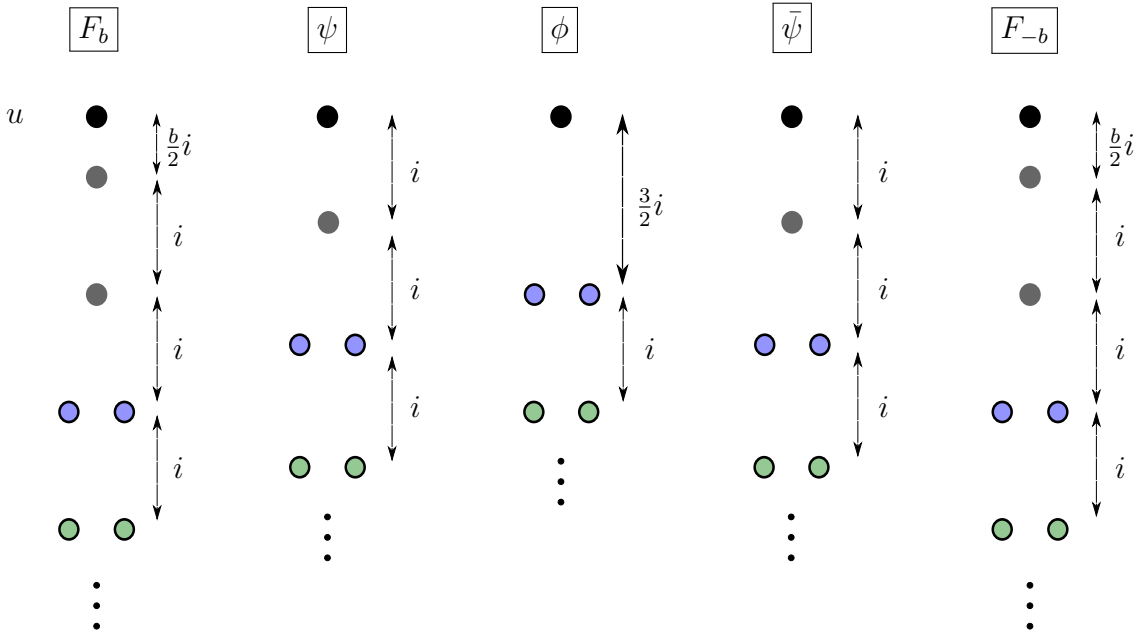

Now we pass to the next descendant $n=2$. If we compute the matrix part as in \eqref{muphimat} we would find the problem that simplifying the sum over residues is not trivial and that the result has a numerator with a one page long polynomial which we omit here. We can avoid this intermediate complication by finding a pattern in which we can take the residues as we did for $n=1$. The integrand of the matrix part is proportional to
\beq
\Pi^{(\text{int})}_{\text{mat}}[\phi(\psi_s\bar\psi_s)^2]\propto\frac{g({\bf w}^1)g({\bf w}^2)g({\bf w}^3)}{f({\bf w}^1,{\bf w}^2)f({\bf w}^2,{\bf w}^3)f({\bf w}^1,{\bf v})f({\bf w}^2,u)f({\bf w}^3,\bf{\bar v})}\,,\la{intmuphimat2}
\eeq
where now each set of small fermions and auxiliary rapidities has two elements. Note that \eqref{intmuphimat2} includes the matrix part integrand for $n=1$. If we start integrating out the first half of the rapidities $\{w^2_1,w^1_1,w^3_1,v_1,\bar v_1\}$ following the pattern derived for $n=1$ we find that the pattern for $n=2$ is 
\begin{center}
\includegraphics{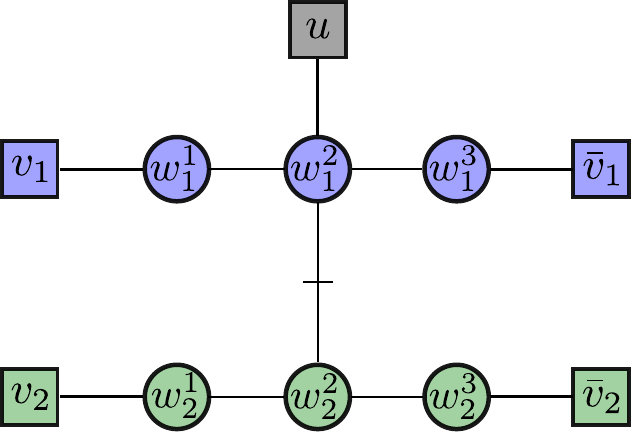}
\end{center}
where the rows in blue and green are separated by $i$. This means that the residues for the second half of the rapidities are evaluated at $\{w^2_2=u-i3/2,w^1_2=u-i2,w^3_2=u-i2,v_2=u-i5/2,\bar v_2=u-i5/2\}$. The Bethe string is then formed by the scalar with rapidity $u$, one small fermion and antifermion at $u-i3/2$ and another pair of small fermion and antifermion at $u-i5/2$. This string is depicted in the third column of figure~\ref{stringsP2P2}.

However, this time we could have used different rapidities provided that they belong to the same set (e.g. take first the residue in $w^2_2$ instead of $w^2_1$). In other words, we can make a permutation of nodes in a given set without altering the outcome. Therefore we need to multiply the result by a combinatoric factor, which in this case is $(2!)^5$ (we have five sets with two rapidities each).

The generalization of these patterns to higher descendants is straightforward. From \eqref{Ks} we see that for each new pair of $(\psi_s\bar\psi_s)$ we have one more auxiliary root of each type. Then all we need to do is to add another row of nodes separated by $i$ to the previous one. From this pattern we can easily see the structure of the Bethe string: the first pair ($n=1$) of small fermion-antifermion attaches to the scalar $3/2i$ below it, for the next pair the separation is $5/2i$ and for the following pairs we keep adding $i$. 

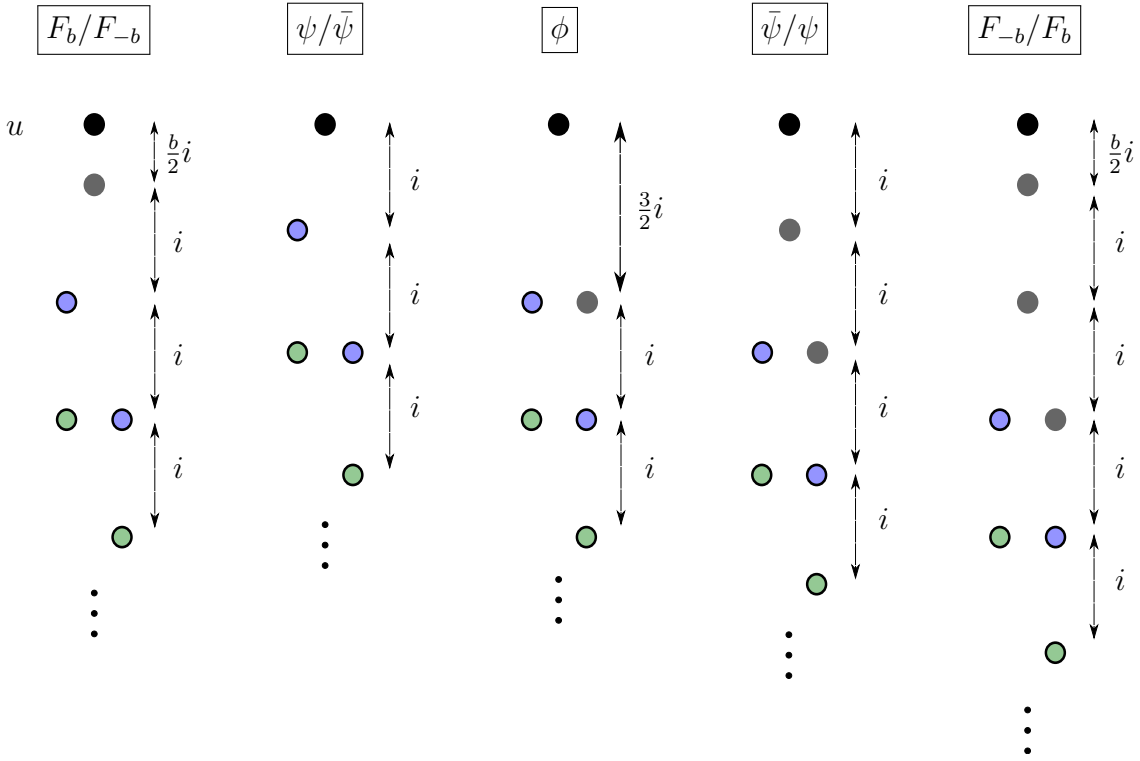
\begin{figure}[ht]
\centering
\def\svgwidth{15cm} \input{stringsP3P1.pdf_tex}
\caption{String patterns in which the small fermions attach to form an effective excitation transforming in the fundamental representation of SU(4). These are the patterns needed for the POPE component $P^{[3]}P^{[1]}$/$P^{[1]}P^{[3]}$. The notation is the same as in Figure~\ref{stringsP2P2}. In this case the fermion and antifermion forming a descendant do not attach at the same distance from the fundamental excitation, but are shifted by $i$.}
\label{stringsP3P1}
\end{figure}

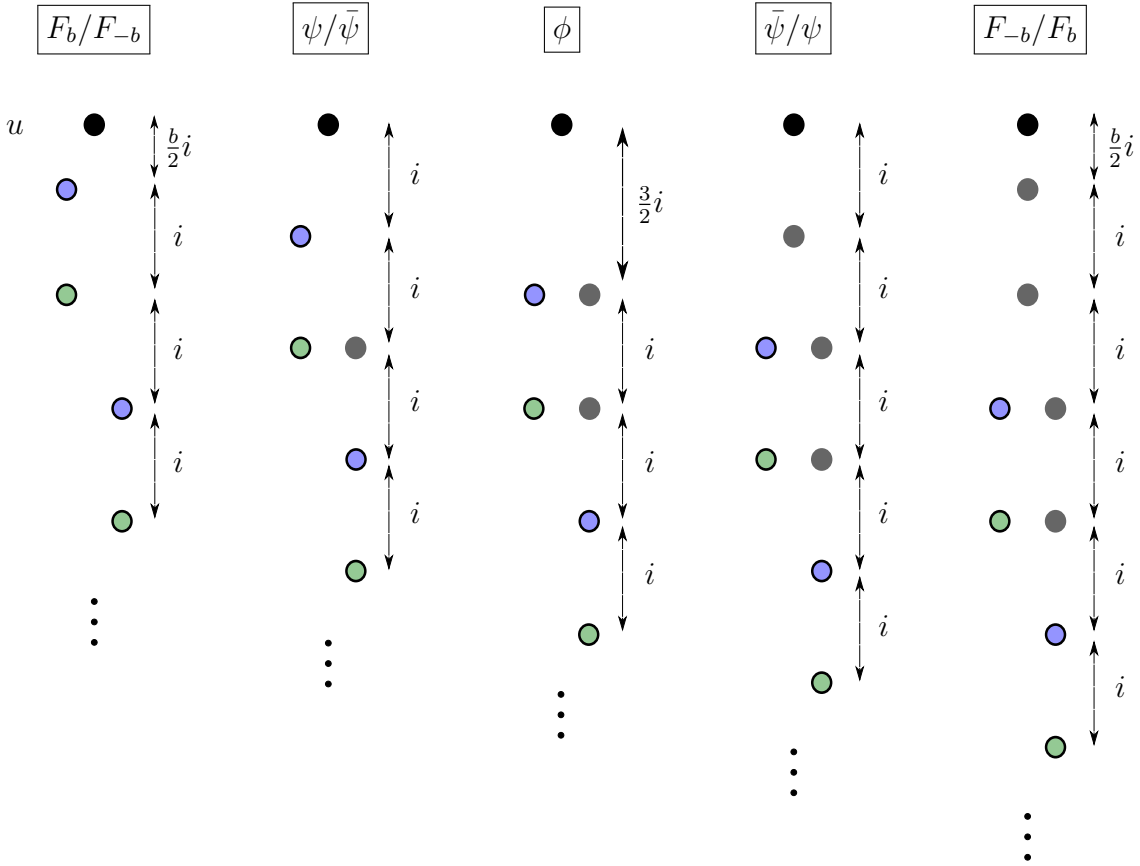
\begin{figure}[ht]
\centering
\def\svgwidth{15cm} \input{stringsP4P0.pdf_tex}
\caption{String patterns for excitations transforming in the singlet representation of SU(4). The notation is the same as in Figure~\ref{stringsP2P2}. The first (second) excitation in each box refers to the pattern appearing at tree level for the POPE component $P^{[4]}P^{[0]}$ ($P^{[0]}P^{[4]}$). For the MHV ($\overline{\text{MHV}}$) case~$P^{[0]}P^{[0]}(P^{[4]}P^{[4]})$ all excitations appear first at one loop.
}
\label{stringsP4P0}
\end{figure}
For other fundamental excitations we can derive similar patterns. In appendix~\ref{app matrix} we explore other examples and give more details on the general structure of these patterns. A general feature is that the separation between small fermions is always $i$, so to know the Bethe strings the only piece of information we need is the separation between the fundamental excitation and the first small fermion and antifermion. That is, if the fundamental excitation has a rapidity $u$, the first small fermion (antifermion) that is attached will be evaluated at a rapidity $u-i\xi_{1(2)}$, where $\xi_{1(2)}$ varies depending on the fundamental excitation. For instance, for a scalar we have $\xi_1=\xi_2=3/2$, for a fermion we would find $\xi_1=1,\,\xi_2=2$, for a positive helicity $b$ gluon bound state $\xi_1=|b|/2,\,\xi_2=|b|/2+2$ and similarly for the conjugate excitations. The strings for excitations in a given representation of the R-symmetry group are presented in figures~\ref{stringsP2P2}-\ref{stringsP4P0}. Notice that the only difference between the different representations is that the small fermions (or antifermions) close to the top can be either part of the primary excitation (gray) or a descendant (blue/green).

Now that we know how the strings of small fermions form we can compute the energy, momentum and angular momentum of the effective excitation. They are simply given by the sum of the individual pieces evaluated at the corresponding rapidities in the Bethe string. For instance, for the state $\phi\psi_s\bar\psi_s$ studied above we obtain $E_{\phi\psi_s\bar\psi_s}(u)=E_{\phi}(u)+E_{\psi_s}(u-3/2i)+E_{\bar\psi_s}(u-3/2i)$. 

Similarly (although the calculation is a bit more involved), we can compute the one effective particle measure $\mu^{[r_1,r_2]}_\Phi(u)$. It has the same universal structure as the one for a fundamental particle and reads
\beq
\mu^{[r_1,r_2]}_\Phi(u)=\[g^{\tfrac{1}{8}r_1(r_1-4)+\tfrac{1}{8}r_2(r_2-4)}\;h_\Phi(u)^{r_2-r_1}\]\;\frac{M_\Phi(u)}{f_{\Phi_0}(u)f_{\Phi_0}(-u)}\;\texttt{exp}_\Phi(u)\la{mu_finitecoupling}\,,
\eeq
where the functions $f_X(u)$ are given in appendix A of \cite{FF} \footnote{For small fermions we have $f_{\psi_s(\bar\psi_s)}(u)=1$, hence only $f_{\Phi_0}$ appears in \eqref{mu_finitecoupling}.}. The exponential part is given by
\beqa
\texttt{exp}_\Phi(u)&=&\text{exp}\[-2\kappa_{\Phi}(u)^t\cdot\mathcal M\cdot\kappa_{\Phi}(u)+2\tilde\kappa_{\Phi}(u)^t\cdot\mathcal M\cdot\tilde\kappa_{\Phi}(u)\]\,,
\eeqa
with
\beq
\kappa_{\Phi}(u)=\kappa_{\Phi_0}(u)+\sum\limits_{k=1}^{N_{\psi_s}}\kappa_{\psi_s}(u-i(\xi_1+k-1))+\sum\limits_{k=1}^{N_{\bar\psi_s}}\kappa_{\bar\psi_s}(u-i(\xi_2+k-1))\,,
\eeq
and similarly for $\tilde\kappa_{\Phi}(u)$, where $\xi_{1(2)}$ label the position at which the first small fermion (antifermion) attaches to the fundamental excitation as above. The vectors $\kappa_X$ and matrix $\mathcal M$ are given in appendix C of \cite{2pt}. The form of this vector is reminiscent of the one for gluons after fusion \cite{fusion}. 

The factor in square brackets in \eqref{mu_finitecoupling} is present only for NMHV amplitudes. The form factor $h_{\Phi}(u)$ can be computed straightforwardly and reads
\beq
h_{\Phi}(u)=h_{\Phi_0}(u)\[\prod\limits_{k=1}^{N_{\psi_s}}h_{\psi_s}(u-i(\xi_1+k-1))\]\[\prod\limits_{k=1}^{N_{\bar\psi_s}}h_{\bar\psi_s}(u-i(\xi_2+k-1))\]\,.
\eeq
Because the form factors satisfy $h_{\bar\Phi}(u)h_{\Phi}(u)=1$, in general there will be many cancellations. For instance, for the component $P^{[2]}P^{[2]}$ all the individual form factors exactly cancel. For other components only some of the first and last small fermions/antifermions contribute to the form factor (this can be seen straightforwardly in Figures~\ref{stringsP2P2}-\ref{stringsP4P0} since there are pairs of fermion-antifermion with the same rapidity).

The prefactor $M_{\Phi}(u)$ is obtained from the product of different prefactors $F_{X|Y}(u,v)$ contained in the pentagon transitions between the various components of the effective particle. The explicit formulas for $M_{\Phi}(u)$ are presented in appendix \ref{app 1effpt}. These are the relevant functions at weak coupling that we shall use in the next section to resum the full series and reproduce the six point tree level amplitude. 

\section{Tree level resummation}\la{tree resum}
In the previous section we found that the small fermions attach to a fundamental excitation in simple patterns that are easy to generalize for any number of pairs $(\psi_s\bar\psi_s)^n$ (see figures~\ref{stringsP2P2}-\ref{stringsP4P0}). With this information we computed the one effective particle measures and form factors. 

As the counting in \cite{FF} shows, the effective one particle states are sufficient to reproduce the full amplitude up to one loop. Here we will focus on the tree level NMHV amplitude, so from now on 
we will assume that the POPE component $P^{[r_1]}P^{[r_2]}$ has $r_1+r_2=4$. As one might expect, several simplifications occur at tree level. Let us look first at the square propagation factor in \eqref{sum1effpt}. The total angular momentum is given by the helicity of the effective particle on the plane. The individual energies can be set to one, so that the total tree level energy is the twist of the effective excitation. Finally, since the small fermion momentum starts at one loop, the total momentum is given by the large excitation which has $p=2u+\mathcal O(g^2)$. To be clear, let us write explicitly these factors for the POPE component $P^{[2]}P^{[2]}$. The relevant excitations transform in the vector representation of SU(4) and are shown in figure~\ref{descendants}. The POPE component reads
\beq
P^{[2]}P^{[2]}=\sum\limits_{n=0}^{\infty}\int\frac{du}{2\pi}e^{i2u\sigma}\[e^{-(1+2n)\tau}\mu^{[2,2]}_{\phi(\psi_s\bar\psi_s)^n}(u)+e^{-(2+2n)\tau+i\phi}\mu^{[2,2]}_{\psi\psi_s(\psi_s\bar\psi_s)^n}(u)+\ldots\]+\cO(g^2)\,,
\eeq
where the measures are evaluated at tree level and the dots represent the contribution of the remaining one effective particle states in the vector representation. 

The effective measures --combined with the corresponding form factors-- are also simplified at tree level. In fact we can easily pack all of them into a single formula where, given the R-charge of the pentagons, we vary the helicity and number of descendants. Here we see explicitly that to describe the possible effective excitations we need to move in the three dimensional space shown in figure~\ref{descendants}. The NMHV measures read
\beq
\mu^{[r_1,r_2]}_{a,n}(u)=\frac{(-1)^{a-\hat r/2} \Gamma \left(\frac{|a|}{2}-i u-\frac{\hat r}{4}\right) \Gamma \left(\frac{|a|}{2}+i u+\frac{3\hat r}{4}\right)}
{\Gamma\(n+1\)\Gamma\(|a|+\frac{\hat r}{2}+n\)}
(-1)^n\(iu+\alpha_+\)_n\(iu+\alpha_-\)_n+\cO(g^2)\,,\la{muan}
\eeq
where $\hat r=r_1(r_2)$ if the excitation has negative (positive) helicity (e.g. for ${[r_1,r_2]=[3,1]}$ we would have $\hat r=1$ for the excitation $\psi$ and $\hat r=3$ for $\phi\bar\psi_s$), $\alpha_{\pm}=1+\tfrac{|a|}{2}+\tfrac{\hat r}{4}\pm\tfrac{|r_{1}-r_2|}{4}$ and $(x)_n$ is the Pochhammer symbol.

Finally, to obtain the tree level NMHV component $P^{[r_1]}P^{[r_2]}$ we simply sum over all possible values of $a$ and $n$.
The result is quite simple and reads
\beq
{P^{[r_1]}P^{[r_2]}}=\delta_{|r_1-r_2|,4}+
\sum\limits_{a,\,n}\int\frac{du}{2\pi}\, e^{-(|a|+\hat r/2+2n)\tau +2iu\sigma + i a\phi}\,\mu^{[r_1,r_2]}_{a,n}+\cO(g^2)\,,\la{Pr1Pr2an}
\eeq
where $\delta_{|r_1-r_2|,4}$ accounts for the vacuum contribution if the excitations allowed are in the singlet representation of $SU(4)$. The sum in $a$ is over the integers or half integers depending on the component we are considering. 

From \eqref{muan} and \eqref{Pr1Pr2an} we can see explicitly how parity symmetry works. Given our definition of $\hat r$, the transformation $r_j\rightarrow 4-r_j$ is equivalent to the replacement $\phi\rightarrow -\phi$ up to an overall sign \footnote{The cases where there is a minus sign can be understood from the exchange on the Grassmann variables $\chi^A$ in the expansion of the superpentagon $\mathbb P$ (see \cite{shortSuper}). For instance, comparing $P^{[3]}P^{[1]}=\mathcal P_{123}\circ\cP_4$ with the parity conjugate of $P^{[1]}P^{[3]}=\cP_1\circ\mathcal P_{234}$ given by $\mathcal P_{234}\circ\cP_1$ we get a minus sign. 
}. This is nothing but the equivalence between NMHV and $\overline{\text{NMHV}}$ for the six point amplitude.

To reproduce the POPE component we now need to perform the sums over $a$ and $n$ and the momentum integral, which is what we turn to next.

\subsection*{Sum over descendants and momentum integral}
The sum over descendants in \eqref{Pr1Pr2an} can be carried out trivially, giving a hypergeometric function
\beqa
&&P^{[r_1]}P^{[r_2]}=\delta_{|r_1-r_2|,4}+\nn\\
&&\sum\limits_{a}\int\frac{du}{2\pi} e^{-(|a|+\hat r/2)\tau +2iu\sigma + i a\phi}\frac{(-1)^{a-\hat r/2} \Gamma \left(\frac{|a|}{2}-i u-\frac{\hat r}{4}\right) \Gamma \left(\frac{|a|}{2}+i u+\frac{3\hat r}{4}\right)}{\Gamma (|a|+\frac{\hat r}{2})}\times\la{2F1}\\
&&\times\,\, {}_2F_1\(\tfrac{|a|}{2}+iu+\tfrac{\hat r}{4}+1-\tfrac{r_{12}}{4},\tfrac{|a|}{2}+iu+\tfrac{\hat r}{4}+1+\tfrac{r_{12}}{4};|a|+\tfrac{\hat r}{2};-e^{-2\tau}\)+\cO(g^2) \nn\,.
\eeqa
This is indeed to be expected since at this perturbative order the $SL(2)$ conformal symmetry is unbroken\footnote{In fact, in \cite{Gaiotto:2011dt} similar expressions were obtained when computing the hexagon remainder function.}. The trick to perform the momentum integral is to trade the sum over descendants $n$ for an integral in a parameter $t$: $\sum_n\rightarrow\int_0^1 dt$, or in other words, use an integral representation for the hypergeometric function. With this replacement all other operations (remaining integrations and sum over helicity) become trivial. The integral representation we shall use is
\beq
{}_2F_1(\textsf{a,b;c;z})=\frac{\Gamma(\textsf c)}{\Gamma(\textsf b)\Gamma(\textsf c-\textsf b )}\int_0^1dt\,t^{\textsf b-1}(1-t)^{\textsf{c-b}-1}(1-t\textsf z)^{-\textsf a}\,.\la{2F1rep}
\eeq
Let us explain how the full procedure works for the component $P^{[2]}P^{[2]}$. After making the replacement \eqref{2F1rep}, we find that the integrand of \eqref{2F1} (with $\hat r=2$ and $r_{12}=0$) takes the simple form
\beq
\int\limits_0^1 dt\, e^{if(t)u}\,g^{[2,2]}(t)\[e^{-\tau-\sigma}(t-1)\]^{|a|}e^{ia\phi}\,,\la{integrandP2P2}
\eeq
where $f(t)=2 \sigma- \ln[(1-t)(1+e^{-2\tau}\, t)/t]$ and $g^{[2,2]}(t)=-t^{1/2}e^{-\tau}(1-t)^{-3/2}(1+te^{-2\tau})^{-3/2}$.  Note that $u$ appears only in the exponent. When the integrand is written in this form, it is apparent that the integration over $u$ trivially gives a delta function $\delta(f(t))$ which in turn localizes $t$ to the value --between zero and one-- where $f(t)=0$. We call this value $t^\star$ and it is given by
\beq
t^\star=\frac{1}{2} e^{2 \tau } \left(\sqrt{\left(e^{2 \sigma }-e^{-2 \tau }+1\right)^2+4 e^{-2 \tau }}-e^{2 \sigma }+e^{-2 \tau }-1\right)\,. 
\eeq
That leaves us with the simple expression
\beq
\int\limits_0^1 dt\int\limits_{-\infty}^\infty\frac{du}{2\pi}\, e^{if(t)u}\,g^{[2,2]}(t)\[e^{-\tau-\sigma}(t-1)\]^{|a|}e^{ia\phi}=
 \frac{g^{[r_1,r_2]}(t^\star)}{|f'(t^\star)|} \[e^{-\tau-\sigma}(t^\star-1)\]^{|a|}e^{ia\phi}\,,\la{I22beforesum}
\eeq
where the prime denotes the derivative of the function with respect to $t$. We might be tempted to sum \eqref{I22beforesum} over  $a$ and equate the result to $P^{[2]}P^{[2]}$, however we need to be a bit more careful. This is because the replacement \eqref{2F1rep} is valid when $\Re(\textsf c)>\Re(\textsf b)>0$, which implies that for some values of the helicity $a$ this replacement is not correct. For the component $P^{[2]}P^{[2]}$ the replacement is strictly valid for $|a|>1$, so we need to perform an analytic continuation.

\begin{figure}
\centering
\includegraphics[scale=1.23]{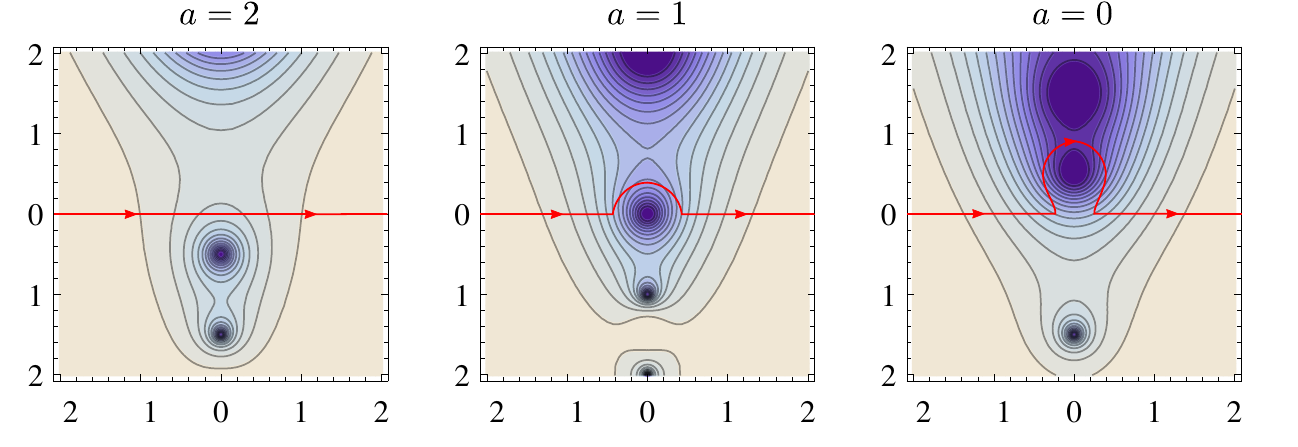}
\caption{Integrand of component $P^{[2]}P^{[2]}$ in the $u$ complex plane for different values of the helicity $a$ (the parameters $\tau$, $\sigma$ and $\phi$ are set to zero); larger values of the integrand are shown in darker colours. In red the contour of integration that gets deformed in performing the analytic continuation of \eqref{I22beforesum}. To the left, the integrand with $a=2$ for which the integral representation is still valid so that the integration contour is over the real line. The first problematic case occurs at $a=1$ (center), where we have a pole at $u=0$; however, we can integrate slightly over the real axis at that point so that effectively the integration contour is unchanged. For $a=0$ (right) we can deform the contour such that we end with the original contour over the real axis minus (clockwise orientation) the residue at $u=i/2$. To get the final result we need to cancel this residue.}
\label{poleP2P2}
\end{figure}
If we analytically continue the result \eqref{I22beforesum} (i.e. integral in $u$) as in figure~\ref{poleP2P2} we see that in deforming the contour of integration we pick an extra term coming from the pole at $u=i/2$ for $a=0$. However, what we want is the analytic continuation of the integrand in $P^{[2]}P^{[2]}$ and then integrate over the real axis. The difference between the two is precisely the residue at $u=i/2$ for $a=0$. Since in the analytic continuation of the integral the contour has clockwise direction the residue comes with a minus sign. That means that in order to get the final result we need to cancel the pole contribution by adding the residue
\beq
r^{[2,2]}=\underset{u=\frac{i}{2}}{\text{Res}}\,\,\bigg[e^{-\tau +2 i  u\sigma }\Gamma \left(-i u-\tfrac{1}{2}\right) \Gamma \left(i u+\tfrac{3}{2}\right) \, _2F_1\left(i u+\tfrac{3}{2},i u+\tfrac{3}{2};1;-e^{-2 \tau }\right)\bigg]=-\frac{e^{-\sigma}}{2\ch\tau}\nn\,.\la{P2P2cont}
\eeq
Although the terms with $|a|=1$ corresponding to $\psi\psi_s(\bar\psi\bar\psi_s)$ and their descendants have a pole at $u=0$ we can simply integrate slightly over the real axis $\mathbb R+i\epsilon$. This is precisely the correct prescription for the integration of the large fermion (see $\cC_{\text{large}}$ in figure~\ref{fermion_contour}).

In the end, the POPE component $P^{[2]}P^{[2]}$ reads
\beq
P^{[2]}P^{[2]}= \frac{g^{[2,2]}(t^\star)}{|f'(t^\star)|}\sum\limits_{a=-\infty}^\infty \[e^{-\tau-\sigma}(t^\star-1)\]^{|a|}e^{ia\phi}+r^{[2,2]}+\cO(g^2)\,.
\eeq
For the remaining components the same procedure applies. In general, we have
\beq
P^{[r_1]}P^{[r_2]}=\delta_{|r_1-r_2|,4}+ \frac{g^{[r_1,r_2]}(t^\star)}{|f'(t^\star)|}\sum\limits_a \[e^{-\tau-\sigma}(t^\star-1)\]^{|a|}e^{ia\phi}+r^{[r_1,r_2]}+\cO(g^2)\,,\la{Pr1Pr2beforesum}
\eeq
where the relevant functions $g^{[r_1,r_2]}(t)$ and $r^{[r_1,r_2]}$ are shown in appendix~\ref{details}.  

Let us emphasize  that the key step in this simplification came from the replacement of the sum over descendants to an integral, which at tree level is straightforward, since it amounts to use one of the integral representations for the hypergeometric function. It remains a question if the same procedure can be easily applied at higher loops\footnote{In \cite{matthotat} the one loop MHV case has been worked out.}. In this way all we are left to do is the last sum over the helicity $a$ which is what we present in the next section.

\subsection*{Sum over helicity}

Finally we can perform the sum over the helicity of the effective excitations. As we see in \eqref{Pr1Pr2beforesum}, the dependence in $a$ is the same for all components and is given by $e^{ia\phi}[e^{-\tau-\sigma}(t^\star-1)]^{|a|}$, so the sum over $a$ is a geometric series. We can perform this sum in the regime where it converges and then analytically continue the result. In particular, the sum converges in the collinear limit (large $\tau$)\footnote{We can see in the definition of $t^\star$ that when taking this limit the dangerous terms outside and inside the square root cancel.}, so we can do the sum close to this region and then analytically continue the result for any value of $\tau$. For example, for $P^{[2]}P^{[2]}$ we do not need to separate the sum and $a$ runs over the integers, so that we find
\beqa
P^{[2]}P^{[2]}&=&\frac{g^{[2,2]}(t^\star)}{|f'(t^*)|}\sum\limits_{a\in\mathbb {Z}} e^{ia\phi}\[e^{-\tau-\sigma}(t^\star-1)\]^{|a|}+r^{[2,2]}\nn\\
&=&\frac{e^{-\tau } \left(e^{\sigma }+2 e^{-\tau } \cos (\phi )\right)}{\left(e^{-2 \tau }+1\right) \left(2 e^{\sigma -\tau } \cos (\phi )+e^{2 \sigma }+e^{-2 \tau }+1\right)}+\cO(g^2)\la{P2P2final}\,,
\eeqa
which nicely matches the data as explained in the next section. Notice that for $P^{[2]}P^{[2]}$ there is a symmetry between positive and negative helicity. In general this is not the case, so when $r_1\neq r_2$ we need to separate the sum for negative and positive values of $a$. Following the same procedure with the other components we complete the resummation of the hexagon POPE series at tree level.

\subsection*{Comparison with tree NMHV amplitude}\la{comparison}

To compare against data we use the map between amplitude and POPE components that was put forward in \cite{shortSuper}. Recall that from supersymmetry \cite{elvang} we need only five NMHV components to express any other hexagon NMHV component. The map between the linear independent components and the POPE basis we have been using is very simple for the hexagon. It is given by
\beqa
P^{[4]}P^{[0]}&=&(({\bf 1})_1)^4\;\;\;\;\;\;\;\;\;\;\;\cW^{(1111)}\,,\nn\\
P^{[3]}P^{[1]}&=&(({\bf 1})_1)^3\,\,({\bf 4})_2\;\;\;\,\cW^{(1114)}\,,\nn\\
P^{[2]}P^{[2]}&=&(({\bf 1})_1)^2(({\bf 4})_2)^2\,\cW^{(1144)}\,,\\
P^{[1]}P^{[3]}&=&\;\,({\bf 1})_1\;\;\,(({\bf 4})_2)^3\,\cW^{(1444)}\,,\nn\\
P^{[0]}P^{[4]}&=&\;\;\;\;\;\;\;\;\;\;\,(({\bf 4})_2)^4\,\cW^{(4444)}\,,\nn\la{mapHEX}
\eeqa
where $\cW$ is the renormalized Wilson loop introduced in \cite{short}, $({\bf i})_j$ denotes the weight of the $i$-th twistor in pentagon $j$ as in \cite{shortSuper} and we have used cyclic labelling for the edges. The hexagon twistors are given in Appendix~\ref{app_twistors}. In this case the weights evaluate to $-1$ for the first, third and last line and $+1$ for the other two. 

At tree level we can compare directly the renormalized Wilson loop $\cW^{\text{NMHV}}$ with the NMHV ratio function of colour-ordered amplitudes $\cR^{\text{NMHV}}=\cA^{\text{NMHV}}/\cA^{\text{MHV}}$~\footnote{We can also write this ratio as $\cR^{\text{NMHV}}=\cW^{\text{NMHV}}/\cW^{\text{MHV}}$. At loop level one would also need to consider the contribution from $\cW^{\text{MHV}}$ in the denominator.}. As can be derived from the recursion relations \cite{Brandhuber:2008pf,ArkaniHamed:2008gz}, the six point NMHV ratio function $\cR_6^{\text{NMHV}}$ at tree level is given by the sum of R-invariants \cite{Drummond:2008vq,Drummond:2008cr}
\beqa
&&\cR_{6,\text{ tree}}^{\text{NMHV}}=R_{135}+R_{136}+R_{146}\,,\text{   where}\\
&&R_{ijk}=\frac{\delta^{(4)}\(\<j-1,j,k-1,k\>\eta_i+\text{cyclic}\)}{\<i,j-1,j,k-1\>\<j-1,j,k-1,k\>\<j,k-1,k,i\>\<k-1,k,i,j-1\>\<k,i,j-1,j\>}\nn\,,
\eeqa
and we have expressed the R-invariants in terms of momentum twistors reviewed in Appendix~\ref{app_twistors}. The delta function ensures that we have a polynomial of degree four in the dual Grassmann variables $\eta_i$. In practice we work with the specific set of $\eta$'s which correspond to a specific component of \eqref{mapHEX}\footnote{Alternatively, we could extract these ratio function components from the package \cite{Bourjaily:2013mma} which computes also one loop ratio functions.}. For example, the component $\cR^{(1144)}$ reads 
\beq
P^{[2]}P^{[2]}=-\cR_{6,\text{ tree}}^{(1144)}=-\frac{\<2345\>\<5123\>}{\<1234\>\<3451\>\<4512\>}-\frac{\<3456\>\<5613\>}{\<1345\>\<4561\>\<6134\>}\la{ratio1144}\,.
\eeq
Finally, in order to compare against the POPE resummed expressions we only need to plug in the twistors in the relevant tree level ratio functions. Doing so for the ratio function component \eqref{ratio1144} we find precisely the tree level term for $P^{[2]}P^{[2]}$ shown in \eqref{P2P2final}. Proceeding in a similar fashion for the rest of the components we find perfect agreement for all of them.

\section{Conclusions}\la{conclusions}
In this paper we presented the tree level resummation of the hexagon POPE reproducing the six point NMHV amplitude. We did so by summing over all possible one effective particle states.

First, we found a way to perform all the small fermion integrals by examining the matrix part of the POPE integrand. We discovered that the small fermions attach to a fundamental excitation following simple patterns creating the strings shown in figures~\ref{stringsP2P2}-\ref{stringsP4P0}. This allowed us to compute the one effective particle measures and form factors at finite coupling. 

The one effective particle states are characterized by their helicity $a$, number of descendants $n$ and $SU(4)$ R-symmetry representation. We found that the NMHV tree level measures $\mu^{[r_1,r_2]}_{a,n}(u)$ can be written in the compact formula \eqref{muan} in terms of these parameters. Given a POPE component, we converted the sum over all possible one effective particle states into a sum over the helicity $a$ and number of descendants $n$, so that a general POPE component has the form
\beq
P^{[r_1]}P^{[r_2]}=\delta_{|r_1-r_2|,4}+\sum_{a,n}\int\frac{du}{2\pi}\;\hat\mu^{[r_1,r_2]}_{a,n}(u)+\cO(g^2)\,.\nn
\eeq

The tree level resummation turned out to be very simple. Once we performed the sums and rapidity integral in the following order
\beq
\sum\limits_{n}\rightarrow\int\frac{du}{2\pi}\rightarrow\sum_a\nn
\eeq
and used some identities for special functions, we recovered the simple rational functions of the tree level six point NMHV amplitudes.

Of course, the ideal case would be to perform the finite coupling resummation. This would make manifest some of the symmetries of the amplitudes --like cyclicity-- obscured in the POPE series. A natural step in that direction is to repeat the procedure described here at higher loops or with larger polygons. In fact, in \cite{matthotat} it is shown that the one loop MHV hexagon can be resummed using the techniques discussed here. Starting from the heptagon Wilson loop, the pentagon transitions between effective excitations will be necessary. Finding methods like \cite{georgiosjames,georgios} to systematically resum all contributions at a given perturbative order would prove most useful. It would also be interesting to find connections between the different looking weak and strong coupling resummations.

The simple patterns found at finite coupling and the almost straightforward resummation of the hexagon at tree level shed an optimistic light on the POPE program as an efficient method for computing the full kinematical regime of scattering amplitudes for larger number of particles and higher loop orders where less is known about them.

\section*{Acknowledgements}
We thank P. Vieira for invaluable discussions, B. Basso and A. Sever for comments on the manuscript and ICTP-SAIFR for hospitality. L.C. is funded by a CONACyT doctoral scholarship. Research at the Perimeter Institute is supported in part by the Government of Canada through NSERC and by the Province of Ontario through MRI.
\appendix
\numberwithin{equation}{section}
\section{More on matrix part and formation of Bethe strings}\la{app matrix}

In this Appendix we give more details on the evaluation of the matrix part of the POPE integrand and the formation of the Bethe strings. Let us start by explaining the overall symmetry factors. For one effective particle states, the symmetry factor $S_n$ in \eqref{gralhex} is given by $S_n=1/(N_{\psi_s}!N_{\bar\psi_s}!)$~\footnote{This is trivial to see when the fundamental excitation is a gluon bound state or a scalar. When the fundamental excitation is a large fermion we have that $S_n=N_{\psi}/(N_{\psi}!N_{\bar\psi_s}!)$, where the numerator counts the possible cases in which fermion is large; since $N_{\psi}=N_{\psi_s}+1$ we have that indeed $S_n=1/(N_{\psi_s}!N_{\bar\psi_s}!)$.}. From the matrix part we also have an overall factor of $1/(K_1!K_2!K_3!)$. As mentioned in the main text, when choosing an order in which we take the residues for the auxiliary and small fermion rapidities $\{{\bf w},{\bf v},{\bf \bar v}\}$, we need to multiply by a combinatoric factor that takes into account all other possible orderings. This factor is simply given by the possible permutations between the different sets of rapidities over which we are integrating: $N_{\psi_s}!N_{\bar\psi_s}!\times K_1!K_2!K_3!$. As we can see, this factor exactly cancels the overall factors previously mentioned. Therefore the effective measure will be given by
\beq
\mu_\Phi(u)=\underset{\{{\bf w},{\bf v},{\bf \bar v}\}=\{{\bf w^*},{\bf v^*},{\bf \bar v^*}\}}{\text{Res}}\[\Pi_{\text{dyn}}\times\Pi_{\text{FF}}\times\Pi^{\text{(int)}}_{\text{mat}}\]
\eeq
where $\Pi^{\text{(int)}}_{\text{mat}}$ is the integrand in \eqref{mat} and $\{{\bf w^*},{\bf v^*},{\bf \bar v^*}\}$ are the positions of the rapidities in the patterns for integration discussed below.
\begin{figure}
\centering
\includegraphics[scale=1]{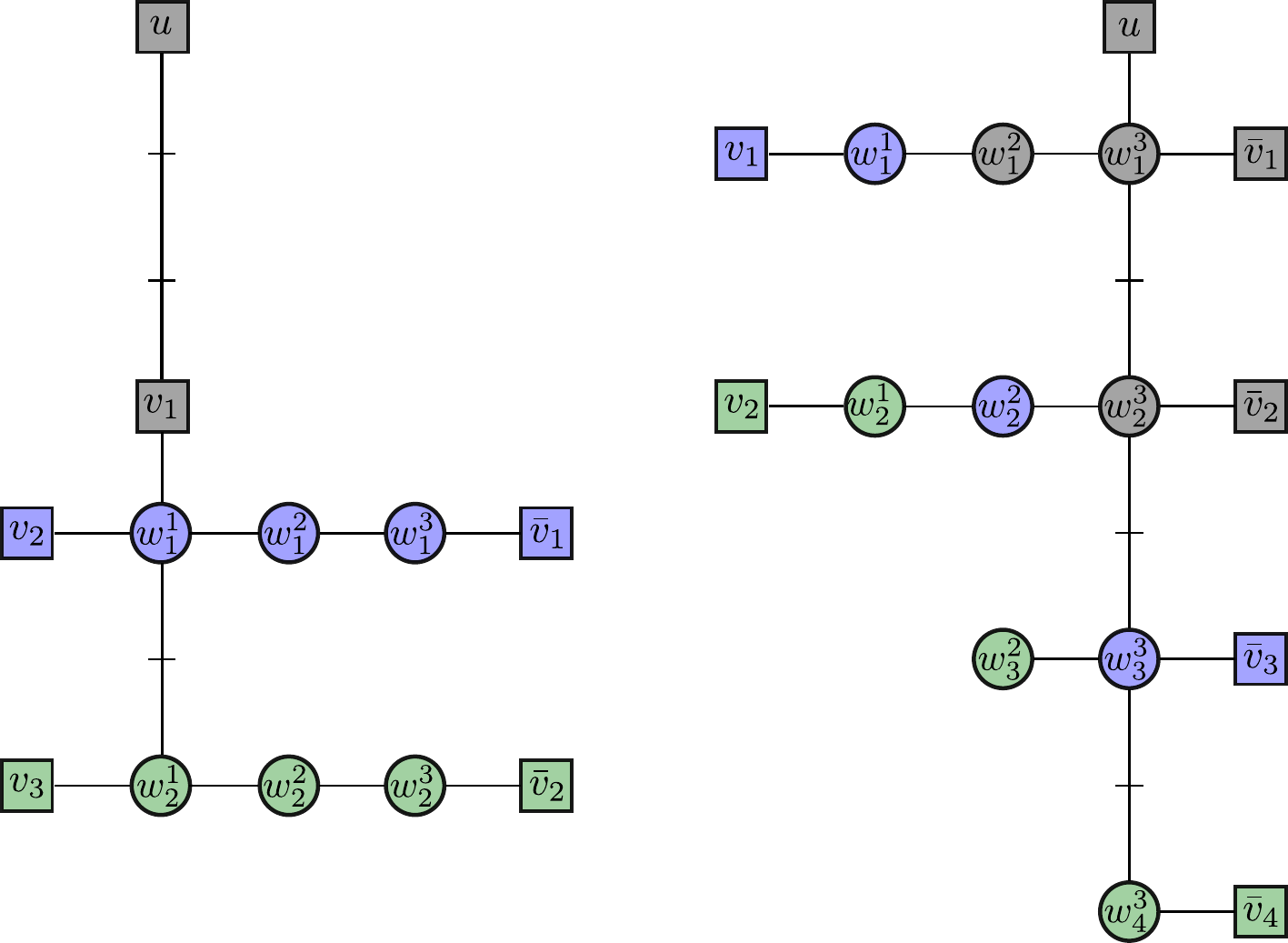}
\caption{Patterns for matrix part integration for the flux tube states $F_3\psi_s(\psi_s\bar\psi_s)^2$ (left) and $\bar\psi\bar\psi_s\bar\psi_s(\psi_s\bar\psi_s)^2$ (right). The top node corresponds to the fundamental excitation with rapidity $u$. The rest of the nodes are integrated out by taking residues at the positions $u-i\#/2$, where $\#$ is the number of line segments between the node we are integrating out and the fundamental one. The nodes forming the primary excitation are coloured in gray and in blue (green) the nodes that are added when we consider the first (second) descendant.}
\la{fig app matrix}
\end{figure}

Now let us see how we can determine the order in which we need to take the residues with two examples. The first one we shall study is the effective measure of the state $F_3\psi_s(\psi_s\bar\psi_s)^n$. Recall that the relevant poles are contained only in the matrix part of the POPE integrand. Since we are redefining the pentagon transitions between gluon bound states and small fermions, we need to include a new function $h^b(u,{\bf v})=\prod_{j}[(u-v_j)^2+(\tfrac{b}{2})^2]$ in the matrix part integrand, which in this case reads 
\beqa
\Pi^{(\text{int})}_{\text{mat}}&=&\frac{g({\bf w}^1)g({\bf w}^2)g({\bf w}^3)}{f({\bf w}^1,{\bf w}^2)f({\bf w}^2,{\bf w}^3)f({\bf w}^1,{\bf v})f({\bf w}^3,{\bf \bar v})h^3( u,{\bf v})}\,.\la{intMATapp}
\eeqa
%The pattern for taking the residues is depicted in figure~\ref{fig app matrix} for $n=2$. 
For $n=0$ the integrand \eqref{intMATapp} reduces to $1/h^3(u,v)$ so that we only need to take the residue at $v_1=u-i3/2$. This is shown in gray in figure~\ref{fig app matrix} with three line segments separating the $u$ and $v$ nodes (for a gluon of positive helicity $b$ there will be $b$ line segments separating the nodes). 

For $n=1$ there is one auxiliary root of each type and only the functions $f$ and $h$ are present in the integrand. Some of the rapidities appear only once in the denominator so that we know immediately which residue we should take. For example, the small antifermion with rapidity $\bar v_1$ appears only in the function $f(w^3_1,\bar v_1)$ so we know that we should take the residue at $\bar v_1=w^3_1-i/2$, afterwhich we are in the same situation for $w^3_1$ and so on until we arrive at the structure in gray and blue in figure~\ref{fig app matrix}. Of course, this is equivalent to take the following sequence of residues: $\{v_1=u-i3/2,\,w^1_1=u-i4/2,\,v_2=u-i5/2,\,w^2_1=u-i5/2,\,w^3_1=u-i6/2,\,\bar v_1=u-i6/2\}$. 

For $n=2$ it is not as straightforward since each variable appears in more than one function so there are several options for which residues to take. The idea is to repeat the pattern found for $n=1$ for each new set of rapidities and unite each row with an effective link. If we integrate all the rapidities  but $w^1_1,w^1_2$ in each row we get an effective pole $1/((w^1_1-w^1_2)^2+1)$ so that --although at the beginning it was prohibited by the function $g(w^1_1,w^1_2)$-- in the end we can take the residue at $w^1_2=w^1_1-i$. The same effective link between different rows is found for higher $n$ and all other excitations. The pattern is summarized in the structure in the left of figure~\ref{fig app matrix}. The order in which we take the poles is important. A simple way to get to right answer is to start from the top of the pattern and take the residues of the nodes closer to the fundamental excitation. This pattern gives rise to the first Bethe string in figure~\ref{stringsP3P1}.
 
Considering other states, the number of auxiliary roots of each type might be different (in \eqref{Ks} we see that it depends exclusively on the representation and excitations of the flux tube state), so each level might not be "complete" as in the previous example. Rather, when we add a descendant we fill the node closer to the fundamental excitation. This is what is depicted the second pattern of integration in figure~\ref{fig app matrix} for the state $\bar\psi\bar\psi_s\bar\psi_s(\psi_s\bar\psi_s)^2$. In figure~\ref{stringsP3P1} it corresponds to the fourth Bethe string. 

As we can see in these examples, the complexity of the patterns of integration increases with the number of excitations in the primary state. However, the maximum number of fermions we can have in a primary state is four, so the most complicated pattern is a slight modification of the example on the right in figure~\ref{fig app matrix}.

Finally, since we are taking all residues in the lower half of the complex plane and the integrations over the auxliary rapidities $w_i$ are over the real line, one has to multiply by an overall factor of $(-1)^{N_w}$, where $N_w$ is the total number of auxiliary rapidities.

In practice, the residues were computed for the first few descendants of all the primary excitations in figure~\ref{tableWEAK}. From them we guessed the pattern for any $n$ resulting in the proposals in the next section.
\section{Measure prefactors at finite coupling}\la{app 1effpt}

In this appendix we present the factors $M_{\Phi}(u)$ in \eqref{mu_finitecoupling} which are the most important functions at leading order in perturbation theory. Since we are dealing with somewhat lengthy equations, let us introduce the notation $[ \pentagon ]=x(u-i\,\pentagon)$ with the usual Zhukowsky variable $x(u)=\tfrac{1}{2}(u+\sqrt{u^2-4g^2})$. The prefactor $M_{\Phi}(u)$ can be conveniently factorized into two pieces, one which contains the contribution from the primary excitations in figure~\ref{tableWEAK} and another one that takes into account its descendants. We identify an excitation by the parameter $\hat r$ which tells us in which $SU(4)$ representation it transforms\footnote{Note that we are keeping $\hat r$ as for the tree level NMHV measures, but in general it does not necessarily correspond to the R-charge of one of the pentagons. For instance, the measures for the MHV component include excitations in the two singlet lines (first and last row in figure~\ref{tableWEAK}) but we keep using $\hat r=4,0$ or $(\hat r=4,0)$ to differentiate between the two.} and the helicity $a$. The prefactors for the effective measures read
\beqa
&&M_\Phi(u)=M_{\Phi_\text{plane}}(u)\;\frac{1}{\Gamma\(n+1\)\Gamma\(|a|+n+\tfrac{\hat r}{2}\)}\times\la{prefactor_finite}\\
&&\prod\limits_{l=1}^n [l_+]^{2-\gb}[l_-]^{\gb}\frac{\sqrt{[l_-]^2-g^2}}{\sqrt{[l_+]^2-g^2}}\;\frac{\([l_+][\ga]-g^2\)\([l_-][1-\ga]-g^2\)\([l_-][l_-+1]-g^2\)}{\([l_-][\ga]-g^2\)\([l_+][1-\ga]-g^2\)\([l_+][l_-+1]-g^2\)}\nn\,.
\eeqa
where $l_\pm=l+\tfrac{|a|}{2}+\tfrac{\hat r}{4}\pm\tfrac{|r_{12}|}{4}$, $\ga=\tfrac{a}{2}+\tfrac{\hat r}{4}$ and
\beq
\gb=\begin{cases}
0\;\;\;\text{singlet,}\nn\\
\frac{1}{2}\;\;\;\text{fundamental/antifundamental,}\\
1\;\;\;\text{vector.}
\end{cases}
\eeq
The constants coming from the matrix part as well as the dynamical part, nicely combine into the two gamma functions shown in the first line of \eqref{prefactor_finite}.

Next we present the prefactors for the primary excitations. For excitations transforming in the vector SU(4) representation we have
\beqa
M_\phi(u)&=&\frac{\pi g}{\cosh(\pi u)}\,,\\
M_{\psi\psi_s}(u)&=&-\frac{\pi g u}{\sinh(\pi u)}\frac{[1]}{[0]}\frac{1}{\sqrt{[0]^2-g^2}\sqrt{[1]^2-g^2}}([0][1]-g^2)\,,\\
M_{F_b\psi_s\psi_s}(u)&=&-g (-1)^b \frac{\Gamma\(iu+\tfrac{|b|}{2}+1\)\Gamma\(-iu+\tfrac{|b|}{2}+1\)}{\Gamma\(|b|\)}\times\\
&&\times\frac{\[\tfrac{|b|}{2}+1\]}{\[-\tfrac{|b|}{2}\]}\frac{1}{\sqrt{\[-\tfrac{|b|}{2}\]^2-g^2}\sqrt{\[\tfrac{|b|}{2}+1\]^2-g^2}}\(\[-\tfrac{|b|}{2}\]\[\tfrac{|b|}{2}+1\]-g^2\)\nn\,.
\eeqa
For particles in the fundamental SU(4) representation
\beqa
M_{F_b\psi_s}(u)&=&ig^{5/4}(-1)^b\frac{\Gamma\(iu+\tfrac{|b|}{2}+1\)\Gamma\(-iu+\tfrac{|b|}{2}+1\)}{\Gamma\(|b|\)}\times\nn\\
&&\times\frac{1}{\[-\tfrac{|b|}{2}\]^{1/2}\sqrt{\[-\tfrac{|b|}{2}^2\]-g^2}}\,,\\
M_{\psi}(u)&=&i\frac{g^{5/4} \pi u}{\sinh(\pi u)}\frac{1}{[0]^{1/2}}\frac{1}{\sqrt{[0]^2-g^2}}\,,\\
M_{\phi\bar\psi_s}(u)&=&-i\frac{g^{5/4} \pi}{\cosh(\pi u)}\[\tfrac{3}{2}\]^{3/2}\frac{1}{\sqrt{\[\tfrac{3}{2}\]^2-g^2}}\,,\\
M_{\bar\psi\bar\psi_s\bar\psi_s}(u)&=&i\frac{g^{5/4} \pi u}{\sinh(\pi u)}\frac{[1]^{3/2}[2]^{3/2}}{[0]^{3/2}}\frac{1}{\sqrt{[0]^2-g^2}\sqrt{[1]^2-g^2}\sqrt{[2]^2-g^2}}\times\nn\\
&&\times\frac{([0][1]-g^2)([0][2]-g^2)}{([1][2]-g^2)}\,,\\
M_{F_{-b}\bar\psi_s\bar\psi_s\bar\psi_s}(u)&=&ig^{5/4}(-1)^b \frac{\Gamma\(iu+\tfrac{|b|}{2}+1\)\Gamma\(iu+\tfrac{|b|}{2}+1\)}{\Gamma\(|b|\)}\frac{\[\tfrac{|b|}{2}+1\]^{3/2}\[\tfrac{|b|}{2}+2\]^{3/2}}{\[-\tfrac{|b|}{2}\]^{3/2}}\times\nn\\
&&\times\frac{1}{\sqrt{\[-\tfrac{|b|}{2}\]^2-g^2}\sqrt{\[\tfrac{|b|}{2}+1\]^2-g^2}\sqrt{\[\tfrac{|b|}{2}+2\]^2-g^2}}\times\\
&&\times\frac{\(\[-\tfrac{|b|}{2}\]\[\tfrac{|b|}{2}+1\]-g^2\)\(\[-\tfrac{|b|}{2}\]\[\tfrac{|b|}{2}+2\]-g^2\)}{\(\[\tfrac{|b|}{2}+1\]\[\tfrac{|b|}{2}+2\]-g^2\)}\nn\,.
\eeqa
The measures for the conjugate excitations in the antifundamental representation are given by the same expressions multiplied by $(-1)$.\\
Finally, for excitations transforming in the singlet SU(4) representation we have
\beqa
M_{F_b}(u)&=&g^2(-1)^b\frac{\Gamma\(iu+\tfrac{|b|}{2}+1\)\Gamma\(-iu+\tfrac{|b|}{2}+1\)}{\Gamma\(|b|\)}\frac{1}{\sqrt{\[-\tfrac{|b|}{2}\]^2-g^2}\sqrt{\[\tfrac{|b|}{2}\]^2-g^2}}\times\nn\\
&&\times\frac{1}{\(\[-\tfrac{|b|}{2}\]\[\tfrac{|b|}{2}\]-g^2\)}\,,\\
M_{\psi\bar\psi_s}(u)&=&\frac{g^2 \pi u}{\sinh(\pi u)}[2]^2\frac{1}{\sqrt{[0]^2-g^2}\sqrt{[2]^2-g^2}}\frac{1}{([0][2]-g^2)}\,,\\
M_{\phi\bar\psi_s\bar\psi_s}(u)&=&-\frac{g^2 \pi}{\cosh(\pi u)}\[\tfrac{3}{2}\]^{2}\[\tfrac{5}{2}\]^{2}\frac{1}{\sqrt{\[\tfrac{3}{2}\]^2-g^2}\sqrt{\[\tfrac{5}{2}\]^2-g^2}}\frac{1}{\(\[\frac{3}{2}\]\[\frac{5}{2}\]-g^2\)}\,,\\
M_{\bar\psi\bar\psi_s\bar\psi_s\bar\psi_s}(u)&=& \frac{g^2 \pi u}{\sinh(\pi u)}\frac{[1]^2[2]^2[3]^2}{[0]^2}\frac{1}{\sqrt{[0]^2-g^2}\sqrt{[1]^2-g^2}\sqrt{[2]^2-g^2}\sqrt{[2]^2-g^2}}\times\nn\\
&&\times\frac{([0][1]-g^2)([0][2]-g^2)([0][3]-g^2)}{([1][2]-g^2)([2][3]-g^2)([3][1]-g^2)}\,,\\
M_{F_{-b}\bar\psi_s\bar\psi_s\bar\psi_s\bar\psi_s}(u)&=&g^2(-1)^b \frac{\Gamma\(iu+\tfrac{|b|}{2}+1\)\Gamma\(iu+\tfrac{|b|}{2}+1\)}{\Gamma\(|b|\)}\frac{\[\tfrac{|b|}{2}+1\]^2\[\tfrac{|b|}{2}+2\]^2\[\tfrac{|b|}{2}+3\]^2}{\[-\tfrac{|b|}{2}\]^2}\times\nn\\
&&\times\frac{1}{\sqrt{\[-\tfrac{|b|}{2}\]^2-g^2}\sqrt{\[\tfrac{|b|}{2}+1\]^2-g^2}\sqrt{\[\tfrac{|b|}{2}+2\]^2-g^2}\sqrt{\[\tfrac{|b|}{2}+3\]^2-g^2}}\times\\
&&\times\frac{\(\[-\tfrac{|b|}{2}\]\[\tfrac{|b|}{2}+1\]-g^2\)\(\[-\tfrac{|b|}{2}\]\[\tfrac{|b|}{2}+2\]-g^2\)\(\[-\tfrac{|b|}{2}\]\[\tfrac{|b|}{2}+3\]-g^2\)}{\(\[\tfrac{|b|}{2}+1\]\[\tfrac{|b|}{2}+2\]-g^2\)\(\[\tfrac{|b|}{2}+2\]\[\tfrac{|b|}{2}+3\]-g^2\)\(\[\tfrac{|b|}{2}+3\]\[\tfrac{|b|}{2}+1\]-g^2\)}\nn\,,
\eeqa
and the same for the conjugate excitations.
\section{Details on momentum integration}\la{details}
In this section we present the relevant functions for the resummation of general POPE components. For clarity, let us rewrite \eqref{Pr1Pr2beforesum} 
%which gives the tree level POPE component after using the integral representation for the hypergeometric function and performing the analytic continuation in $a$
\beq
P^{[r_1]}P^{[r_2]}=\delta_{|r_1-r_2|,4}+ \frac{g^{[r_1,r_2]}(t^\star)}{|f'(t^\star)|}\sum\limits_a \[e^{-\tau-\sigma}(t-1)\]^{|a|}e^{ia\phi}+r^{[r_1,r_2]}+\cO(g^2)\,,
\eeq
where $f(t)=2 \sigma- \ln[(1-t)(1+e^{-2\tau}\, t)/t]$ and the other functions depend on the POPE component we are considering.

Although for the case studied in the main text there was a symmetry between positive and negative helicity states, this is in general not the case. This can be seen from \eqref{2F1} where the integrand depends explicitly on $\hat r$. This means that for each component we have two different functions depending on the value of $\hat r$
\beqa
g^{[4,0]([0,4])}&=&\begin{cases}
\dfrac{t^2e^{-2\tau}}{(1-t)^2(1+te^{-2\tau})}\,,\,\;\;\;\;\;\;\;\;\;\;\;\;\;\;\,\hat r=4,\nn\\[5mm]
\dfrac{1}{t(1-t)(1+te^{-2\tau})^2}\,,\,\;\;\;\;\;\;\;\;\;\;\;\;\;\,\hat r=0\,.
\end{cases}\\
g^{[3,1]([1,3])}&=&\begin{cases}
\dfrac{t^{5/4}e^{-3\tau/2}}{(1-t)^{7/4}(1+te^{-2\tau})^{5/4}}\,,\;\;\;\;\;\;\;\;\;\hat r=3,\\[5mm]
-\dfrac{e^{-\tau/2}}{t^{1/4}(1-t)^{5/4}(1+te^{-2\tau})^{7/4}}\,,\;\hat r=1\,.
\end{cases}\\
g^{[2,2]}&=&-\dfrac{t^{1/2}e^{-\tau}}{(1-t)^{3/2}(1+te^{-2\tau})^{3/2}}\,,\,\;\;\;\;\;\;\;\;\;\hat r=2,\nn
\eeqa
The other relevant functions $r^{[r_1,r_2]}$ arise when performing the analytical continuation in $a$.
For the component $P^{[4]}P^{[0]}(P^{[0]}P^{[4]})$ the function comes from taking the residue of the integrand in \eqref{2F1} at $u=i$ with $a=0$ and at $u=i/2$ with $a=-1(+1)$
\beqa
r^{[4,0]}(\sigma,\tau)&=&e^{-2 (\sigma +\tau )} \left(-1+\frac{e^{\sigma +\tau -i \phi }}{e^{2 \tau }+1}\right)\,,\\
r^{[0,4]}(\sigma,\tau)&=&e^{-2 (\sigma +\tau )} \left(-1+\frac{e^{\sigma +\tau +i \phi }}{e^{2 \tau }+1}\right)\,.
\eeqa
For $P^{[3]}P^{[1]}(P^{[1]}P^{[3]})$ the residues are taken at $u=i$ with $a=-\tfrac{1}{2}(+\tfrac{1}{2})$ and at $u=i$ with $a=\tfrac{1}{2}(-\tfrac{1}{2})$ and the functions read
\beqa
r^{[3,1]}(\sigma,\tau)&=&e^{-2 \sigma -\tau -\frac{i \phi }{2}} \left(-\frac{e^{\sigma -\tau }}{e^{-2 \tau }+1}+e^{i \phi }\right)\,,\\
r^{[1,3]}(\sigma,\tau)&=&-e^{-2 \sigma -\tau +\frac{i \phi }{2}} \left(-\frac{e^{\sigma -\tau }}{e^{-2 \tau }+1}+e^{-i \phi }\right)\,.
\eeqa
Note that these functions are the same (up to a sign) after we make the replacement $\phi\rightarrow-\phi$. For completeness we rewrite $r^{[2,2]}$ which is found by taking the residue at $u=i/2$ with $a=0$
\beq
r^{[2,2]}(\sigma,\tau)=\frac{1}{2} e^{-\sigma } \text{sech}(\tau )\,.
\eeq

\section{Hexagon twistors}\la{app_twistors}
In this appendix we review how the kinematical data enters in the POPE approach. We shall use momentum twistors which are very useful variables since they trivialize momentum conservation and on-shellness. They are four components vectors (spinors of $\mathbb R^{2,4}$) defined up to rescaling $Z\simeq tZ$. Each twistor $Z_i$ is associated to an edge $i$ of the polygon. For example, for the hexagon we have\\
\beq
\begin{minipage}{.35\textwidth}
\def\svgwidth{3cm} 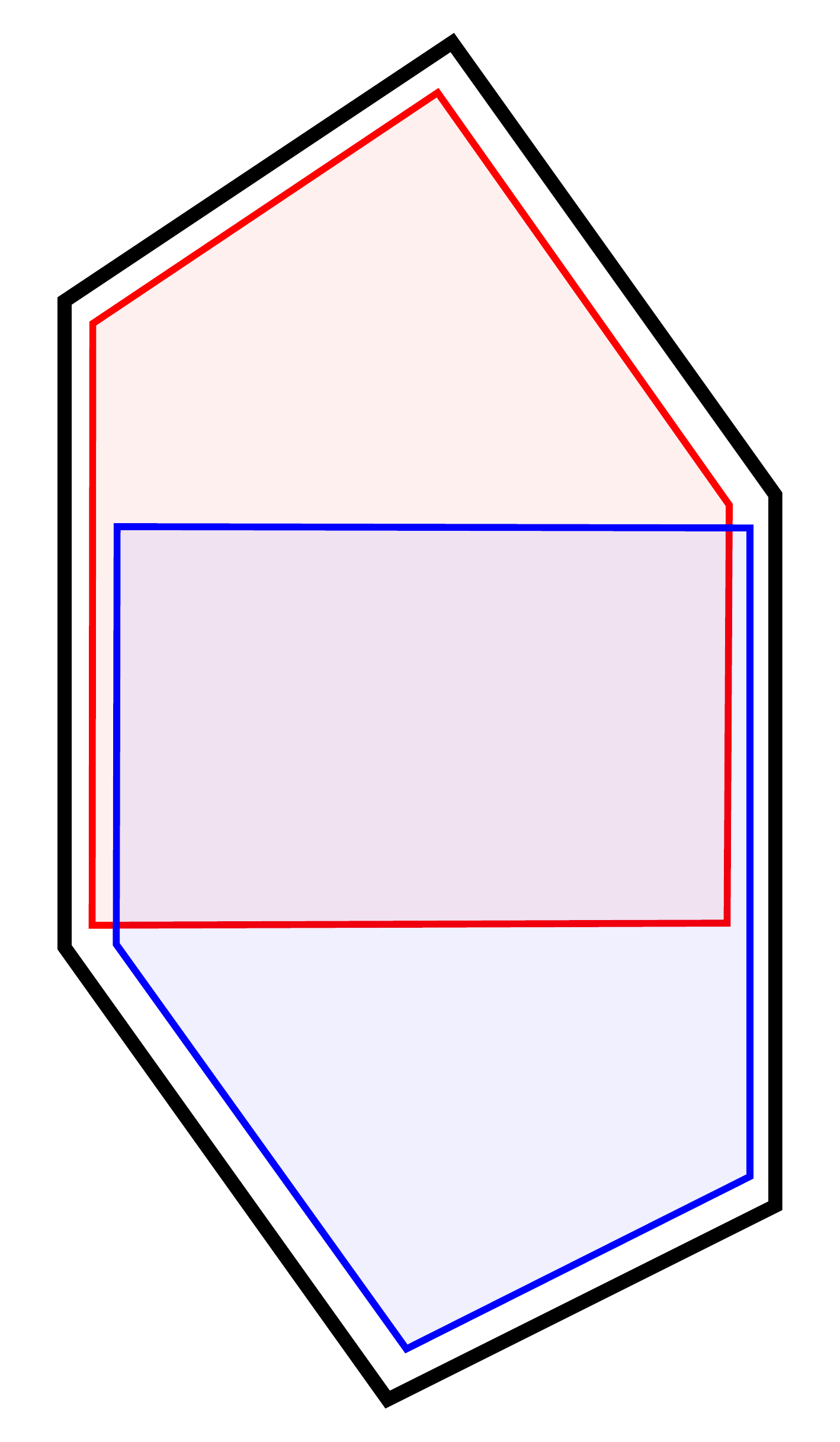
\end{minipage}
\left(
\begin{array}{c}
 Z_1\\
 Z_2\\
 Z_3\\
 Z_4\\
 Z_5\\
 Z_6
\end{array}
\right)
=
\left(
\begin{array}{cccc}
 e^{\sigma -\frac{i \phi }{2}} & 0 & e^{\tau +\frac{i \phi }{2}} & e^{\frac{i \phi }{2}-\tau } \\
 e^{\sigma -\frac{i \phi }{2}} 
 %1
 & 0 & 0 & 0 \\
 -1 & 0 & 0 & 1 \\
 0 & 1 & -1 & 1 \\
 0 & 1 & 0 & 0 \\
 0 & e^{-\sigma -\frac{i \phi }{2}} & e^{\tau +\frac{i \phi }{2}} & 0 \\
\end{array}
\right)
\eeq

The hexagon twistors are constructed by acting with the symmetries of the middle square on the bottom of the polygon as explained in appendix A of \cite{data}. The variables $\tau$, $\sigma$ and $\phi$ parametrize the three conformal symmetries that the middle square preserves \cite{Alday:2010ku} and play the role of flux tube time, space and angle coordinates. They can be related to the three cross ratios $\{u_1,\,u_2,\,u_3\}$ of the hexagon (see again \cite{data} for the explicit relations).

As mentioned in the main text, the tree level and one loop NMHV components can be extracted from the package \cite{Bourjaily:2013mma}. For instance, to extract the component $P^{[2]}P^{[2]}$ at tree level and evaluate with the above twistors, we simply write
\begin{center}
\verb|evaluate@superComponent[{1,2},{},{},{3,4},{},{}]@treeAmp[6,1]|
\end{center}
and multiply the result by the appropriate weights defined in \cite{shortSuper} which in this case combine to $\pm1$.

\end{document}

%% file: stringsP2P2.pdf_tex
\begingroup%
  \makeatletter%
  \providecommand\color[2][]{%
    \errmessage{(Inkscape) Color is used for the text in Inkscape, but the package 'color.sty' is not loaded}%
    \renewcommand\color[2][]{}%
  }%
  \providecommand\transparent[1]{%
    \errmessage{(Inkscape) Transparency is used (non-zero) for the text in Inkscape, but the package 'transparent.sty' is not loaded}%
    \renewcommand\transparent[1]{}%
  }%
  \providecommand\rotatebox[2]{#2}%
  \ifx\svgwidth\undefined%
    \setlength{\unitlength}{1008.71386719bp}%
    \ifx\svgscale\undefined%
      \relax%
    \else%
      \setlength{\unitlength}{\unitlength * \real{\svgscale}}%
    \fi%
  \else%
    \setlength{\unitlength}{\svgwidth}%
  \fi%
  \global\let\svgwidth\undefined%
  \global\let\svgscale\undefined%
  \makeatother%
  \begin{picture}(1,.7)%
    \put(0,0){\includegraphics[width=\unitlength]{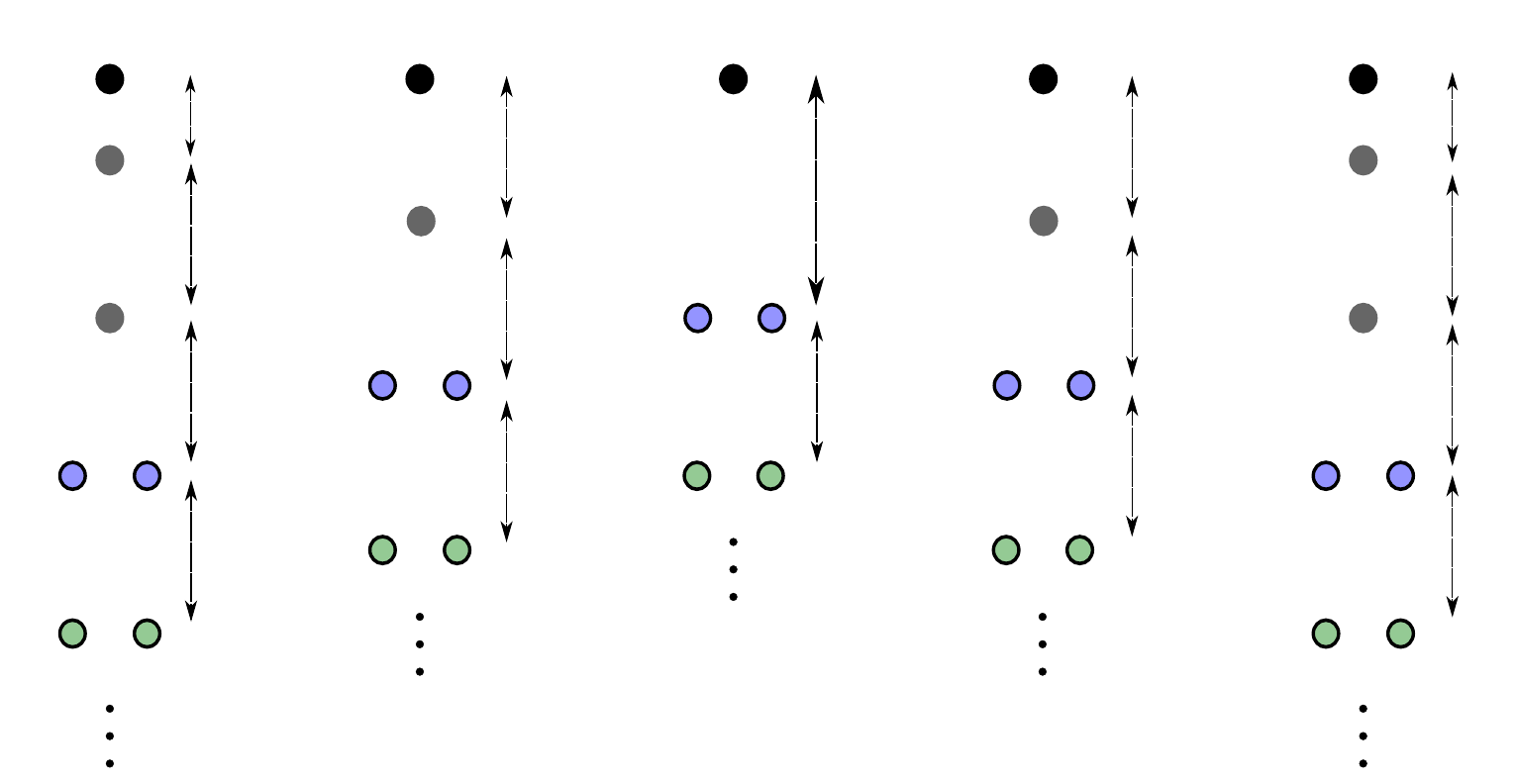}}%

    \put(0.483,0.65){\color[rgb]{0,0,0}\makebox(0,0)[cc]{\smash{\scalebox{1.3}{\textsf{vector}
    %P$^{[2]}$P$^{[2]}$
    }}}}%      
    
    \put(0.07,0.535){\color[rgb]{0,0,0}\makebox(0,0)[cc]{\smash{\fbox{$F_b$}}}}%
    \put(0.277,0.535){\color[rgb]{0,0,0}\makebox(0,0)[cc]{\smash{\fbox{$\psi$}}}}%
    \put(0.483,0.535){\color[rgb]{0,0,0}\makebox(0,0)[cc]{\smash{\fbox{$\phi$}}}}%
    \put(0.689,0.535){\color[rgb]{0,0,0}\makebox(0,0)[cc]{\smash{\fbox{$\bar\psi$}}}}%
    \put(0.895,0.535){\color[rgb]{0,0,0}\makebox(0,0)[cc]{\smash{\fbox{$F_{-b}$}}}}%
   
    \put(0.0,0.4599){\color[rgb]{0,0,0}\makebox(0,0)[cc]{\smash{$u$}}}%
   
    \put(0.145,0.431){\color[rgb]{0,0,0}\makebox(0,0)[cc]{\smash{$\tfrac{b}{2}i$}}}%
    \put(0.145,0.35){\color[rgb]{0,0,0}\makebox(0,0)[cc]{\smash{$i$}}}%
    \put(0.145,0.25){\color[rgb]{0,0,0}\makebox(0,0)[cc]{\smash{$i$}}}%
    \put(0.145,0.145){\color[rgb]{0,0,0}\makebox(0,0)[cc]{\smash{$i$}}}%    
    
    \put(0.355,0.407){\color[rgb]{0,0,0}\makebox(0,0)[cc]{\smash{$i$}}}%
    \put(0.355,0.303){\color[rgb]{0,0,0}\makebox(0,0)[cc]{\smash{$i$}}}%
    \put(0.355,0.199){\color[rgb]{0,0,0}\makebox(0,0)[cc]{\smash{$i$}}}%
    
    \put(0.563,0.378){\color[rgb]{0,0,0}\makebox(0,0)[cc]{\smash{$\tfrac{3}{2}i$}}}%
    \put(0.563,0.25){\color[rgb]{0,0,0}\makebox(0,0)[cc]{\smash{$i$}}}%
    
    \put(0.77,0.407){\color[rgb]{0,0,0}\makebox(0,0)[cc]{\smash{$i$}}}%
    \put(0.77,0.303){\color[rgb]{0,0,0}\makebox(0,0)[cc]{\smash{$i$}}}%
    \put(0.77,0.199){\color[rgb]{0,0,0}\makebox(0,0)[cc]{\smash{$i$}}}%

    \put(0.98,0.431){\color[rgb]{0,0,0}\makebox(0,0)[cc]{\smash{$\tfrac{b}{2}i$}}}%
    \put(0.98,0.35){\color[rgb]{0,0,0}\makebox(0,0)[cc]{\smash{$i$}}}%
    \put(0.98,0.25){\color[rgb]{0,0,0}\makebox(0,0)[cc]{\smash{$i$}}}%
    \put(0.98,0.145){\color[rgb]{0,0,0}\makebox(0,0)[cc]{\smash{$i$}}}%     
  \end{picture}%
\endgroup%

%% file: stringsP3P1.pdf_tex
\begingroup%
  \makeatletter%
  \providecommand\color[2][]{%
    \errmessage{(Inkscape) Color is used for the text in Inkscape, but the package 'color.sty' is not loaded}%
    \renewcommand\color[2][]{}%
  }%
  \providecommand\transparent[1]{%
    \errmessage{(Inkscape) Transparency is used (non-zero) for the text in Inkscape, but the package 'transparent.sty' is not loaded}%
    \renewcommand\transparent[1]{}%
  }%
  \providecommand\rotatebox[2]{#2}%
  \ifx\svgwidth\undefined%
    \setlength{\unitlength}{1008.71386719bp}%
    \ifx\svgscale\undefined%
      \relax%
    \else%
      \setlength{\unitlength}{\unitlength * \real{\svgscale}}%
    \fi%
  \else%
    \setlength{\unitlength}{\svgwidth}%
  \fi%
  \global\let\svgwidth\undefined%
  \global\let\svgscale\undefined%
  \makeatother%
  \begin{picture}(1,.8)%
    \put(0,0){\includegraphics[width=\unitlength]{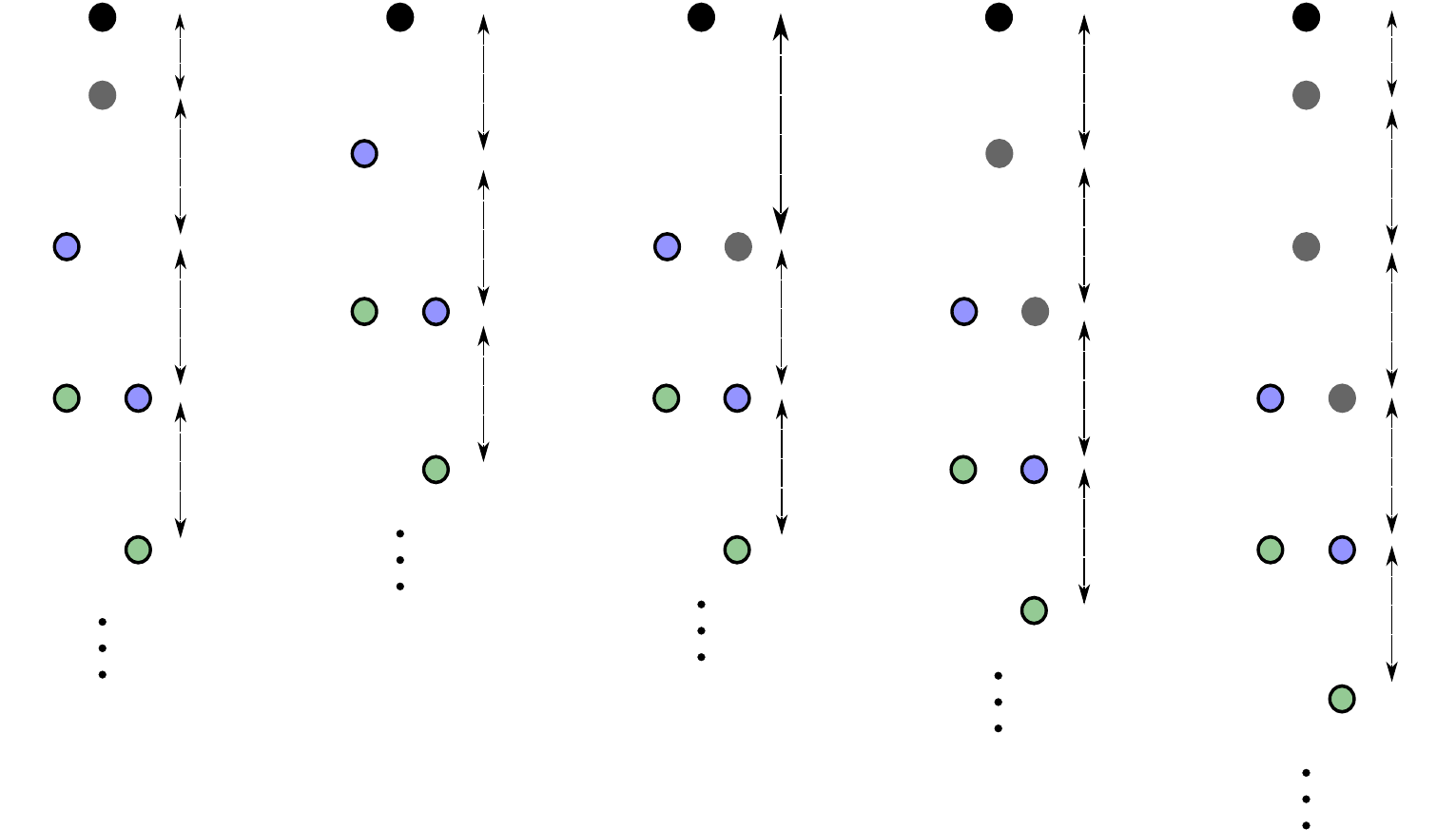}}%
    
    \put(0.483,0.75){\color[rgb]{0,0,0}\makebox(0,0)[cc]{\smash{\scalebox{1.3}{\textsf{fundamental/antifundamental}
    %P$^{[3]}$P$^{[1]}$
    }}}}%    
    
    \put(0.07,0.635){\color[rgb]{0,0,0}\makebox(0,0)[cc]{\smash{\fbox{$F_b/F_{-b}$}}}}%
    \put(0.275,0.635){\color[rgb]{0,0,0}\makebox(0,0)[cc]{\smash{\fbox{$\psi/\bar\psi$}}}}%
    \put(0.483,0.635){\color[rgb]{0,0,0}\makebox(0,0)[cc]{\smash{\fbox{$\phi$}}}}%
    \put(0.687,0.635){\color[rgb]{0,0,0}\makebox(0,0)[cc]{\smash{\fbox{$\bar\psi/\psi$}}}}%
    \put(0.895,0.635){\color[rgb]{0,0,0}\makebox(0,0)[cc]{\smash{\fbox{$F_{-b}/F_b$}}}}%
   
    \put(0.0,0.549){\color[rgb]{0,0,0}\makebox(0,0)[cc]{\smash{$u$}}}%
    
    \put(0.145,0.525){\color[rgb]{0,0,0}\makebox(0,0)[cc]{\smash{$\tfrac{b}{2}i$}}}%
    \put(0.145,0.445){\color[rgb]{0,0,0}\makebox(0,0)[cc]{\smash{$i$}}}%
    \put(0.145,0.345){\color[rgb]{0,0,0}\makebox(0,0)[cc]{\smash{$i$}}}%
    \put(0.145,0.245){\color[rgb]{0,0,0}\makebox(0,0)[cc]{\smash{$i$}}}%    
    
    \put(0.355,0.5){\color[rgb]{0,0,0}\makebox(0,0)[cc]{\smash{$i$}}}%
    \put(0.355,0.399){\color[rgb]{0,0,0}\makebox(0,0)[cc]{\smash{$i$}}}%
    \put(0.355,0.299){\color[rgb]{0,0,0}\makebox(0,0)[cc]{\smash{$i$}}}%
    
    \put(0.563,0.475){\color[rgb]{0,0,0}\makebox(0,0)[cc]{\smash{$\tfrac{3}{2}i$}}}%
    \put(0.563,0.345){\color[rgb]{0,0,0}\makebox(0,0)[cc]{\smash{$i$}}}%
    \put(0.563,0.245){\color[rgb]{0,0,0}\makebox(0,0)[cc]{\smash{$i$}}}%
    
    \put(0.77,0.5){\color[rgb]{0,0,0}\makebox(0,0)[cc]{\smash{$i$}}}%
    \put(0.77,0.399){\color[rgb]{0,0,0}\makebox(0,0)[cc]{\smash{$i$}}}%
    \put(0.77,0.299){\color[rgb]{0,0,0}\makebox(0,0)[cc]{\smash{$i$}}}%
    \put(0.77,0.199){\color[rgb]{0,0,0}\makebox(0,0)[cc]{\smash{$i$}}}%

    \put(0.98,0.525){\color[rgb]{0,0,0}\makebox(0,0)[cc]{\smash{$\tfrac{b}{2}i$}}}%
    \put(0.98,0.445){\color[rgb]{0,0,0}\makebox(0,0)[cc]{\smash{$i$}}}%
    \put(0.98,0.345){\color[rgb]{0,0,0}\makebox(0,0)[cc]{\smash{$i$}}}%
    \put(0.98,0.245){\color[rgb]{0,0,0}\makebox(0,0)[cc]{\smash{$i$}}}%    
    \put(0.98,0.145){\color[rgb]{0,0,0}\makebox(0,0)[cc]{\smash{$i$}}}%   
  \end{picture}%
\endgroup%

%% file: stringsP4P0.pdf_tex
\begingroup%
  \makeatletter%
  \providecommand\color[2][]{%
    \errmessage{(Inkscape) Color is used for the text in Inkscape, but the package 'color.sty' is not loaded}%
    \renewcommand\color[2][]{}%
  }%
  \providecommand\transparent[1]{%
    \errmessage{(Inkscape) Transparency is used (non-zero) for the text in Inkscape, but the package 'transparent.sty' is not loaded}%
    \renewcommand\transparent[1]{}%
  }%
  \providecommand\rotatebox[2]{#2}%
  \ifx\svgwidth\undefined%
    \setlength{\unitlength}{1008.71386719bp}%
    \ifx\svgscale\undefined%
      \relax%
    \else%
      \setlength{\unitlength}{\unitlength * \real{\svgscale}}%
    \fi%
  \else%
    \setlength{\unitlength}{\svgwidth}%
  \fi%
  \global\let\svgwidth\undefined%
  \global\let\svgscale\undefined%
  \makeatother%
  \begin{picture}(1,.9)%
    \put(0,0){\includegraphics[width=\unitlength]{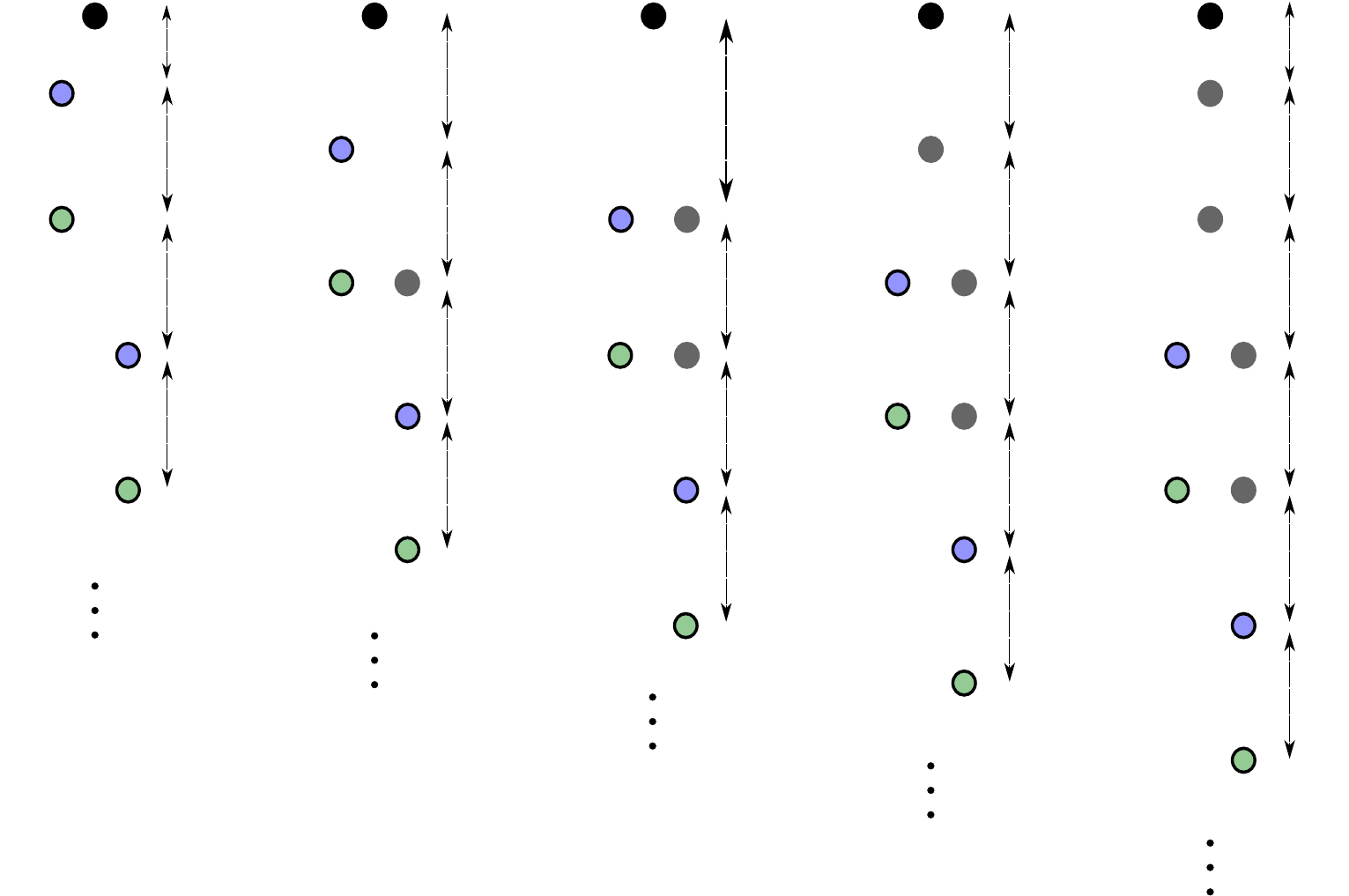}}%
    
    \put(0.485,0.85){\color[rgb]{0,0,0}\makebox(0,0)[cc]{\smash{\scalebox{1.3}{\textsf{singlet}
    %P$^{[4]}$P$^{[0]}$
    }}}}%  
        
    \put(0.07,0.73){\color[rgb]{0,0,0}\makebox(0,0)[cc]{\smash{\fbox{$F_b/F_{-b}$}}}}%
    \put(0.28,0.73){\color[rgb]{0,0,0}\makebox(0,0)[cc]{\smash{\fbox{$\psi/\bar\psi$}}}}%
    \put(0.485,0.73){\color[rgb]{0,0,0}\makebox(0,0)[cc]{\smash{\fbox{$\phi$}}}}%
    \put(0.69,0.73){\color[rgb]{0,0,0}\makebox(0,0)[cc]{\smash{\fbox{$\bar\psi/\psi$}}}}%
    \put(0.90,0.73){\color[rgb]{0,0,0}\makebox(0,0)[cc]{\smash{\fbox{$F_{-b}/F_b$}}}}%
   
    \put(0.0,0.6449){\color[rgb]{0,0,0}\makebox(0,0)[cc]{\smash{$u$}}}%
    
    \put(0.145,0.625){\color[rgb]{0,0,0}\makebox(0,0)[cc]{\smash{$\tfrac{b}{2}i$}}}%
    \put(0.145,0.545){\color[rgb]{0,0,0}\makebox(0,0)[cc]{\smash{$i$}}}%
    \put(0.145,0.445){\color[rgb]{0,0,0}\makebox(0,0)[cc]{\smash{$i$}}}%
    \put(0.145,0.345){\color[rgb]{0,0,0}\makebox(0,0)[cc]{\smash{$i$}}}%    
    
    \put(0.355,0.6){\color[rgb]{0,0,0}\makebox(0,0)[cc]{\smash{$i$}}}%
    \put(0.355,0.499){\color[rgb]{0,0,0}\makebox(0,0)[cc]{\smash{$i$}}}%
    \put(0.355,0.399){\color[rgb]{0,0,0}\makebox(0,0)[cc]{\smash{$i$}}}%
    \put(0.355,0.299){\color[rgb]{0,0,0}\makebox(0,0)[cc]{\smash{$i$}}}%
    
    \put(0.563,0.575){\color[rgb]{0,0,0}\makebox(0,0)[cc]{\smash{$\tfrac{3}{2}i$}}}%
    \put(0.563,0.445){\color[rgb]{0,0,0}\makebox(0,0)[cc]{\smash{$i$}}}%
    \put(0.563,0.345){\color[rgb]{0,0,0}\makebox(0,0)[cc]{\smash{$i$}}}%
    \put(0.563,0.245){\color[rgb]{0,0,0}\makebox(0,0)[cc]{\smash{$i$}}}%
    
    \put(0.77,0.6){\color[rgb]{0,0,0}\makebox(0,0)[cc]{\smash{$i$}}}%
    \put(0.77,0.499){\color[rgb]{0,0,0}\makebox(0,0)[cc]{\smash{$i$}}}%
    \put(0.77,0.399){\color[rgb]{0,0,0}\makebox(0,0)[cc]{\smash{$i$}}}%
    \put(0.77,0.299){\color[rgb]{0,0,0}\makebox(0,0)[cc]{\smash{$i$}}}%
    \put(0.77,0.199){\color[rgb]{0,0,0}\makebox(0,0)[cc]{\smash{$i$}}}%

    \put(0.98,0.625){\color[rgb]{0,0,0}\makebox(0,0)[cc]{\smash{$\tfrac{b}{2}i$}}}%
    \put(0.98,0.545){\color[rgb]{0,0,0}\makebox(0,0)[cc]{\smash{$i$}}}%
    \put(0.98,0.445){\color[rgb]{0,0,0}\makebox(0,0)[cc]{\smash{$i$}}}%
    \put(0.98,0.345){\color[rgb]{0,0,0}\makebox(0,0)[cc]{\smash{$i$}}}%    
    \put(0.98,0.245){\color[rgb]{0,0,0}\makebox(0,0)[cc]{\smash{$i$}}}% 
    \put(0.98,0.145){\color[rgb]{0,0,0}\makebox(0,0)[cc]{\smash{$i$}}}% 
  \end{picture}%
\endgroup%

%% file: hexagon.pdf_tex
\begingroup%
  \makeatletter%
  \providecommand\color[2][]{%
    \errmessage{(Inkscape) Color is used for the text in Inkscape, but the package 'color.sty' is not loaded}%
    \renewcommand\color[2][]{}%
  }%
  \providecommand\transparent[1]{%
    \errmessage{(Inkscape) Transparency is used (non-zero) for the text in Inkscape, but the package 'transparent.sty' is not loaded}%
    \renewcommand\transparent[1]{}%
  }%
  \providecommand\rotatebox[2]{#2}%
  \ifx\svgwidth\undefined%
    \setlength{\unitlength}{3025.25449219bp}%
    \ifx\svgscale\undefined%
      \relax%
    \else%
      \setlength{\unitlength}{\unitlength * \real{\svgscale}}%
    \fi%
  \else%
    \setlength{\unitlength}{\svgwidth}%
  \fi%
  \global\let\svgwidth\undefined%
  \global\let\svgscale\undefined%
  \makeatother%
  \begin{picture}(1,1.6)%
    \put(0,0){\includegraphics[width=\unitlength]{hexagon.pdf}}%
    
    \put(0.7639937,0.0657315){\color[rgb]{0,0,0}\makebox(0,0)[cc]{\smash{$1$}}}%
    \put(0.19713192,0.21376216){\color[rgb]{0,0,0}\makebox(0,0)[cc]{\smash{$6$}}}%
    \put(0.011391877,0.93826717){\color[rgb]{0,0,0}\makebox(0,0)[cc]{\smash{$5$}}}%
    \put(1,0.6898654){\color[rgb]{0,0,0}\makebox(0,0)[cc]{\smash{$2$}}}%
    \put(0.25,1.5393431){\color[rgb]{0,0,0}\makebox(0,0)[cc]{\smash{$4$}}}%
    \put(0.78,1.4070162){\color[rgb]{0,0,0}\makebox(0,0)[cc]{\smash{$3$}}}%
    
  \end{picture}%
\endgroup%

%% file: resumhex_final.bbl
\begin{thebibliography}{99}



\bibitem{AM}
  L.~F.~Alday, J.~M.~Maldacena,
  ``Gluon scattering amplitudes at strong coupling,''
  JHEP {\bf 0706}, (2007) 064
  [arXiv:0705.0303].


\bibitem{AmplitudeWilson}
 G.~P.~Korchemsky, J.~M.~Drummond, E.~Sokatchev,
 ``Conformal properties of four-gluon planar amplitudes and Wilson loops,''
  Nucl.\ Phys.\  {\bf B795}, (2008) 385-408
  [arXiv:0707.0243]
$\bullet$  A.~Brandhuber, P.~Heslop, G.~Travaglini,
 ``MHV amplitudes in N=4 super Yang-Mills and Wilson loops,''
  Nucl.\ Phys.\  {\bf B794}, (2008) 231-243
  [arXiv:0707.1153]
  $\bullet$  Z.~Bern, L.~J.~Dixon, D.~A.~Kosower, R.~Roiban, M.~Spradlin, C.~Vergu and A.~Volovich,
``The Two-Loop Six-Gluon MHV Amplitude in Maximally Supersymmetric Yang-Mills Theory,''
  Phys.\ Rev.\  D {\bf 78}, (2008) 045007
  [arXiv:0803.1465]
$\bullet$  J.~M.~Drummond, J.~Henn, G.~P.~Korchemsky and E.~Sokatchev,
``Hexagon Wilson loop = six-gluon MHV amplitude,''
  Nucl.\ Phys.\  B {\bf 815} (2009) 142
  [arXiv:0803.1466]
$\bullet$  N.~Berkovits, J.~Maldacena,
``Fermionic T-Duality, Dual Superconformal Symmetry, and the Amplitude/Wilson Loop Connection,''
  JHEP {\bf 0809}, (2008) 062 
  [arXiv:0807.3196].


\bibitem{short}
  B.~Basso, A.~Sever and P.~Vieira,
  ``Spacetime and Flux Tube S-Matrices at Finite Coupling for N=4 Supersymmetric Yang-Mills Theory,''
  Phys.\ Rev.\ Lett.\  {\bf 111} (2013) 9,  091602
  [arXiv:1303.1396 [hep-th]].
  %%CITATION = ARXIV:1303.1396;%%
  

\bibitem{data}
  B.~Basso, A.~Sever and P.~Vieira,
  ``Space-time S-matrix and Flux tube S-matrix II. Extracting and Matching Data,''
  JHEP {\bf 1401} (2014) 008
  [arXiv:1306.2058 [hep-th]].
  %%CITATION = ARXIV:1306.2058;%%
  

\bibitem{2pt}
  B.~Basso, A.~Sever and P.~Vieira,
  ``Space-time S-matrix and Flux-tube S-matrix III. The two-particle contributions,''
  arXiv:1402.3307 [hep-th].
  %%CITATION = ARXIV:1402.3307;%%


\bibitem{fusion} 
  B.~Basso, A.~Sever and P.~Vieira,
  ``Space-time S-matrix and Flux-tube S-matrix IV. Gluons and Fusion,''
  JHEP {\bf 1409}, 149 (2014)
  [arXiv:1407.1736 [hep-th]].
  %%CITATION = ARXIV:1407.1736;%%


\bibitem{Belitsky:2014sla}
  A.~V.~Belitsky,
  ``Nonsinglet pentagons and NMHV amplitudes,''
  Nucl.\ Phys.\ B {\bf 896} (2015) 493
  [arXiv:1407.2853 [hep-th]].
  %%CITATION = ARXIV:1407.2853;%%


\bibitem{Andrei1}
  A.~V.~Belitsky,
  ``Fermionic pentagons and NMHV hexagon,''
  Nucl.\ Phys.\ B {\bf 894} (2015) 108
  [arXiv:1410.2534 [hep-th]].
  

\bibitem{Andrei2} 
  A.~V.~Belitsky,
  ``On factorization of multiparticle pentagons,''
  arXiv:1501.06860 [hep-th].
  %%CITATION = ARXIV:1501.06860;%%

\bibitem{FrankToAppear}
   B.~Basso, F.~Coronado, A.~Sever and P.~Vieira,
 ``Spacetime and Flux Tube S-Matrices. The Matrix Part'', to appear.


\bibitem{shortSuper}
  B.~Basso, J.~Caetano, L.~Cordova, A.~Sever and P.~Vieira,
  ``OPE for all Helicity Amplitudes,''
  arXiv:1412.1132 [hep-th].
  %%CITATION = ARXIV:1412.1132;%%

\bibitem{FF}
  B.~Basso, J.~Caetano, L.~Cordova, A.~Sever and P.~Vieira,
  ``OPE for all Helicity Amplitudes II. Form Factors and Data analysis,''
  JHEP {\bf 1512} (2015) 088
  doi:10.1007/JHEP12(2015)088
  [arXiv:1508.02987 [hep-th]].
  %%CITATION = doi:10.1007/JHEP12(2015)088;%%


\bibitem{shortHexagon}
  B.~Basso, A.~Sever and P.~Vieira,
  ``Hexagonal Wilson Loops in Planar $\mathcal{N}=4$ SYM Theory at Finite Coupling,''
  arXiv:1508.03045 [hep-th].
  %%CITATION = ARXIV:1508.03045;%%

\bibitem{georgiosjames}
 J.~M.~Drummond and G.~Papathanasiou,
 ``Hexagon OPE Resummation and Multi-Regge Kinematics,''
 arXiv:1507.08982 [hep-th].

\bibitem{georgios}
  G.~Papathanasiou,
  ``Hexagon Wilson Loop OPE and Harmonic Polylogarithms,''
  JHEP {\bf 1311} (2013) 150
  [arXiv:1310.5735 [hep-th]].


\bibitem{Fioravanti:2015dma}
  D.~Fioravanti, S.~Piscaglia and M.~Rossi,
  ``Asymptotic Bethe Ansatz on the GKP vacuum as a defect spin chain: scattering, particles and minimal area Wilson loops,''
  Nucl.\ Phys.\ B {\bf 898} (2015) 301
  doi:10.1016/j.nuclphysb.2015.07.007
  [arXiv:1503.08795 [hep-th]].
  
\bibitem{Bonini:2015lfr}
  A.~Bonini, D.~Fioravanti, S.~Piscaglia and M.~Rossi,
  ``Strong Wilson polygons from the lodge of free and bound mesons,''
  arXiv:1511.05851 [hep-th].
  %%%%%%%%%%%%%%%%%%%%%%%%%
%
%\bibitem{Belitsky:2014rba}
%  A.~V.~Belitsky, S.~E.~Derkachov and A.~N.~Manashov,
%  ``Quantum mechanics of null polygonal Wilson loops,''
%  Nucl.\ Phys.\ B {\bf 882} (2014) 303
%  [arXiv:1401.7307 [hep-th]].
%  %%CITATION = ARXIV:1401.7307;%%
  

\bibitem{AldayMaldacena}
 L.~F.~Alday and J.~M.~Maldacena,
``Comments on operators with large spin,''
 JHEP {\bf 0711} (2007) 019
[arXiv:0708.0672].
%%CITATION = ARXIV:0708.0672;

%
%\bibitem{MoreDispPaper}
% B.~Basso and A.~V.~Belitsky,
%``Luescher formula for GKP string,''
%Nucl.\ Phys.\ B {\bf 860} (2012) 1
%[arXiv:1108.0999].
%%%CITATION = ARXIV:1108.0999;%%  %%CITATION = ARXIV:1010.5237;%%

\bibitem{Gaiotto:2010fk}
  D.~Gaiotto, J.~Maldacena, A.~Sever and P.~Vieira,
  ``Bootstrapping Null Polygon Wilson Loops,''
  JHEP {\bf 1103} (2011) 092
  doi:10.1007/JHEP03(2011)092
  [arXiv:1010.5009 [hep-th]].
  %%CITATION = doi:10.1007/JHEP03(2011)092;%%  
  
\bibitem{Gaiotto:2011dt}
  D.~Gaiotto, J.~Maldacena, A.~Sever and P.~Vieira,
  ``Pulling the straps of polygons,''
  JHEP {\bf 1112} (2011) 011
  doi:10.1007/JHEP12(2011)011
  [arXiv:1102.0062 [hep-th]].
  %%CITATION = doi:10.1007/JHEP12(2011)011;%%
%
%\bibitem{Sever:2012qp}
%  A.~Sever, P.~Vieira and T.~Wang,
%  ``From Polygon Wilson Loops to Spin Chains and Back,''
%  JHEP {\bf 1212} (2012) 065
%  doi:10.1007/JHEP12(2012)065
%  [arXiv:1208.0841 [hep-th]].
%  %%CITATION = doi:10.1007/JHEP12(2012)065;%%  
%  
%\bibitem{Sever:2011pc}
%  A.~Sever and P.~Vieira,
%  ``Multichannel Conformal Blocks for Polygon Wilson Loops,''
%  JHEP {\bf 1201} (2012) 070
%  doi:10.1007/JHEP01(2012)070
%  [arXiv:1105.5748 [hep-th]].
%  %%CITATION = doi:10.1007/JHEP01(2012)070;%%  
%  
\bibitem{Ben}
  B.~Basso,
  ``Exciting the GKP string at any coupling,''
  Nucl.\ Phys.\ B {\bf 857} (2012) 254
  doi:10.1016/j.nuclphysb.2011.12.010
  [arXiv:1010.5237 [hep-th]].
  %%CITATION = doi:10.1016/j.nuclphysb.2011.12.010;%%  
%   
\bibitem{matthotat}
 M. von Hippel and H.T. Lam, in preparation.

\bibitem{elvang}
  H.~Elvang, D.~Z.~Freedman and M.~Kiermaier,
  ``Solution to the Ward Identities for Superamplitudes,''
  JHEP {\bf 1010} (2010) 103
  doi:10.1007/JHEP10(2010)103
  [arXiv:0911.3169 [hep-th]].
  %%CITATION = doi:10.1007/JHEP10(2010)103;%% 
 
%\bibitem{Drummond:2008bq}
%  J.~M.~Drummond, J.~Henn, G.~P.~Korchemsky and E.~Sokatchev,
%  ``Generalized unitarity for N=4 super-amplitudes,''
%  Nucl.\ Phys.\ B {\bf 869} (2013) 452
%  doi:10.1016/j.nuclphysb.2012.12.009
%  [arXiv:0808.0491 [hep-th]].
%  %%CITATION = doi:10.1016/j.nuclphysb.2012.12.009;%% 

\bibitem{Brandhuber:2008pf}
  A.~Brandhuber, P.~Heslop and G.~Travaglini,
  ``A Note on dual superconformal symmetry of the N=4 super Yang-Mills S-matrix,''
  Phys.\ Rev.\ D {\bf 78} (2008) 125005
  doi:10.1103/PhysRevD.78.125005
  [arXiv:0807.4097 [hep-th]].

\bibitem{ArkaniHamed:2008gz}
  N.~Arkani-Hamed, F.~Cachazo and J.~Kaplan,
  ``What is the Simplest Quantum Field Theory?,''
  JHEP {\bf 1009} (2010) 016
  doi:10.1007/JHEP09(2010)016
  [arXiv:0808.1446 [hep-th]].
  

\bibitem{Drummond:2008vq}
  J.~M.~Drummond, J.~Henn, G.~P.~Korchemsky and E.~Sokatchev,
  ``Dual superconformal symmetry of scattering amplitudes in N=4 super-Yang-Mills theory,''
  Nucl.\ Phys.\ B {\bf 828} (2010) 317
  doi:10.1016/j.nuclphysb.2009.11.022
  [arXiv:0807.1095 [hep-th]].
  %%CITATION = doi:10.1016/j.nuclphysb.2009.11.022;%% 
   
  
\bibitem{Drummond:2008cr}
  J.~M.~Drummond and J.~M.~Henn,
  ``All tree-level amplitudes in N=4 SYM,''
  JHEP {\bf 0904} (2009) 018
  doi:10.1088/1126-6708/2009/04/018
  [arXiv:0808.2475 [hep-th]].
  %%CITATION = doi:10.1088/1126-6708/2009/04/018;%% 
 
 
\bibitem{Bourjaily:2013mma} 
  J.~L.~Bourjaily, S.~Caron-Huot and J.~Trnka,
 ``Dual-Conformal Regularization of Infrared Loop Divergences and the Chiral Box Expansion,''
  arXiv:1303.4734 [hep-th].
  %%CITATION = ARXIV:1303.4734;%%
  
\bibitem{Alday:2010ku}
  L.~F.~Alday, D.~Gaiotto, J.~Maldacena, A.~Sever and P.~Vieira,
  ``An Operator Product Expansion for Polygonal null Wilson Loops,''
  JHEP {\bf 1104} (2011) 088
  doi:10.1007/JHEP04(2011)088
  [arXiv:1006.2788 [hep-th]].
  %%CITATION = doi:10.1007/JHEP04(2011)088;%%  
  
%
%\bibitem{superloopsimon}
% S.~Caron-Huot,
% ``Notes on the scattering amplitude / Wilson loop duality,''
% [arXiv:1010.1167].
%
%\bibitem{Korchemsky:2010ut}
%  G.~P.~Korchemsky and E.~Sokatchev,
%  ``Superconformal invariants for scattering amplitudes in N=4 SYM theory,''
%  Nucl.\ Phys.\ B {\bf 839} (2010) 377
%  [arXiv:1002.4625 [hep-th]].
%
%
%\bibitem{heptagonB}
%  J.~M.~Drummond, G.~Papathanasiou and M.~Spradlin,
%  ``A Symbol of Uniqueness: The Cluster Bootstrap for the 3-Loop MHV Heptagon,''
%  JHEP {\bf 1503} (2015) 072
%  [arXiv:1412.3763 [hep-th]].
%  %%CITATION = ARXIV:1412.3763;%%
%  
%\bibitem{LanceEtAl1}
%L.~J.~Dixon, J.~M.~Drummond and J.~M.~Henn,
%``Bootstrapping the three-loop hexagon,''
%JHEP {\bf 1111} (2011) 023
%%[arXiv:1108.4461]. 
%
%\bibitem{LanceEtAl2}
%L.~J.~Dixon, J.~M.~Drummond, M.~von Hippel and J.~Pennington,
% ``Hexagon functions and the three-loop remainder function,''
%JHEP {\bf 1312}, 049 (2013)
%% arXiv:1308.2276 [hep-th].
%
%\bibitem{LanceEtAl3}
%  L.~J.~Dixon, J.~M.~Drummond, C.~Duhr and J.~Pennington,
%  ``The four-loop remainder function and multi-Regge behavior at NNLLA in planar N=4 super-Yang-Mills theory,''
%  arXiv:1402.3300 [hep-th]. 
%  
%\bibitem{LanceEtAl4}
% L.~J.~Dixon, J.~M.~Drummond, C.~Duhr, M.~von Hippel and J.~Pennington,
%  ``Bootstrapping six-gluon scattering in planar N=4 super-Yang-Mills theory,''
%  PoS LL {\bf 2014} (2014) 077
%  [arXiv:1407.4724 [hep-th]].
%  
%\bibitem{LanceEtAl5}
%  L.~J.~Dixon and M.~von Hippel,
%  ``Bootstrapping an NMHV amplitude through three loops,''
%  JHEP {\bf 1410} (2014) 65
%  [arXiv:1408.1505 [hep-th]].
%  

\end{thebibliography}
